\documentclass[prl,aps,twocolumn,superscriptaddress,showpacs]{revtex4}
\usepackage{epsfig}
\usepackage{amsmath}
\usepackage{amsfonts}
\usepackage{amssymb,mathrsfs}
\usepackage{bm}
\usepackage{bbm}

\newcommand{\bs}[1]{\boldsymbol{#1}}

\begin{document}

\title{Coulomb drag in graphene near the Dirac point}

\author{M. Sch\"utt}
\affiliation{Institut f\"ur Nanotechnologie, Karlsruhe Institute of Technology,
 76021 Karlsruhe, Germany}

\author{P.M. Ostrovsky} 
\affiliation{Max-Planck-Institut f\"ur Festk\"orperforschung,
  Heisenbergstr. 1, 70569, Stuttgart, Germany}
\affiliation{Institut f\"ur Nanotechnologie, Karlsruhe Institute of
  Technology, 76021 Karlsruhe, Germany}
\affiliation{L.D. Landau Institute for Theoretical Physics RAS, 119334
  Moscow, Russia}

\author{M. Titov} 
\affiliation{Radboud University Nijmegen,
  Institute for Molecules and Materials, NL-6525 AJ Nijmegen, The
  Netherlands}

\author{I.V. Gornyi} 
\affiliation{Institut f\"ur Nanotechnologie, Karlsruhe Institute of
  Technology, 76021 Karlsruhe, Germany}
\affiliation{A.F. Ioffe Physico-Technical Institute, 194021
  St. Petersburg, Russia}

\author{B.N. Narozhny} 
\affiliation{Institut f\"ur Theorie der Kondensierten Materie and DFG
  Center for Functional Nanostructures, Karlsruher Institut f\"ur
  Technologie, 76128 Karlsruhe, Germany}

\author{A.D. Mirlin}
\affiliation{Institut f\"ur Nanotechnologie, Karlsruhe Institute of Technology,
 76021 Karlsruhe, Germany}
\affiliation{Institut f\"ur Theorie der Kondensierten Materie and DFG
  Center for Functional Nanostructures, Karlsruher Institut f\"ur
  Technologie, 76128 Karlsruhe, Germany}
\affiliation{Petersburg Nuclear Physics Institute,
 188350 St. Petersburg, Russia}

\date{\today}

\begin{abstract}
  We study Coulomb drag in double-layer graphene near the Dirac
  point. A particular emphasis is put on the case of clean graphene,
  with transport properties dominated by the electron-electron
  interaction. Using the quantum kinetic equation framework, we show
  that the drag becomes $T$-independent in the clean limit, $T\tau \to
  \infty$, where $T$ is temperature and $1/\tau$ impurity scattering
  rate. For stronger disorder (or lower temperature), $T\tau \ll
  1/\alpha^2$, where $\alpha$ is the interaction strength, the kinetic
  equation agrees with the leading-order ($\alpha^2$) perturbative
  result. At still lower temperatures, $T\tau \ll 1$ (diffusive
  regime) this contribution gets suppressed, while the next-order
  ($\alpha^3$) contribution becomes important; it yields a peak
  centered at the Dirac point with a magnitude that grows with
  lowering $T\tau$.
\end{abstract}


\maketitle

Frictional drag in double-layer systems consisting of two closely
spaced, but electronically isolated conductors is a well established
experimental tool for studying the microscopic structure of solids
\cite{roj,ex1,ex2,ex3,ex4,ex5,tut}. In such an experiment a current
$I_1$ is passed through one of the conductors (the ``active'' layer)
and the induced voltage drop $V_2$ is measured along the other
(``passive'') layer. The ratio of this voltage to the driving current
$\rho_D=-V_2/I_1$ (known as the drag coefficient or the
transresistivity) is a measure of both the inter-layer interaction
\cite{roj,ex1} and the microscopic state \cite{ex2,ex3,ex4,ex5} of the
layers. At low temperatures the drag effect is dominated by direct
Coulomb interaction between the carriers in the two layers.

\begin{figure}
$
\begin{array}{ccc}
\hspace*{-0.6cm}\epsfig{file=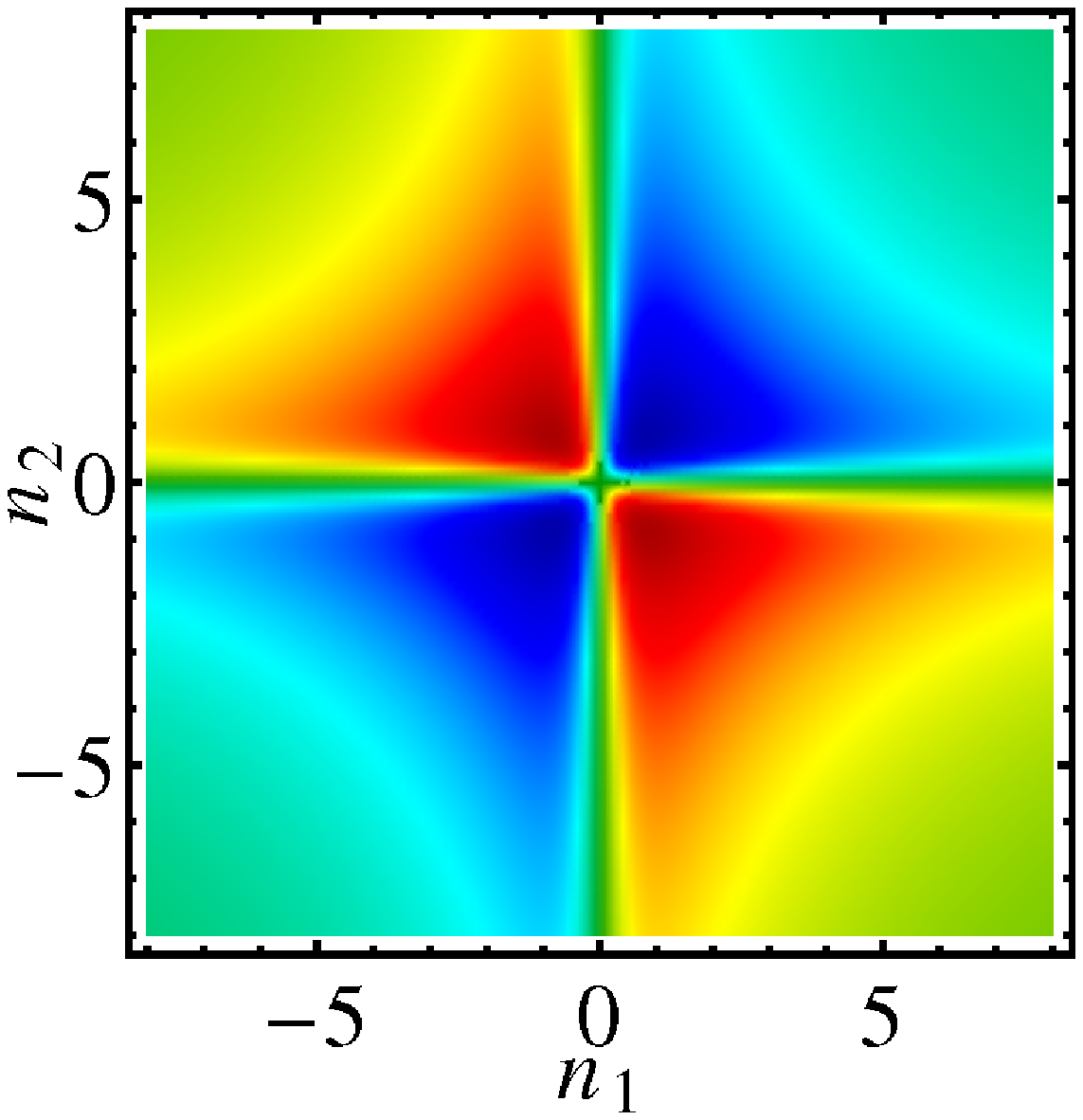,width=3.5cm} &
\hspace*{-0.6cm}\epsfig{file=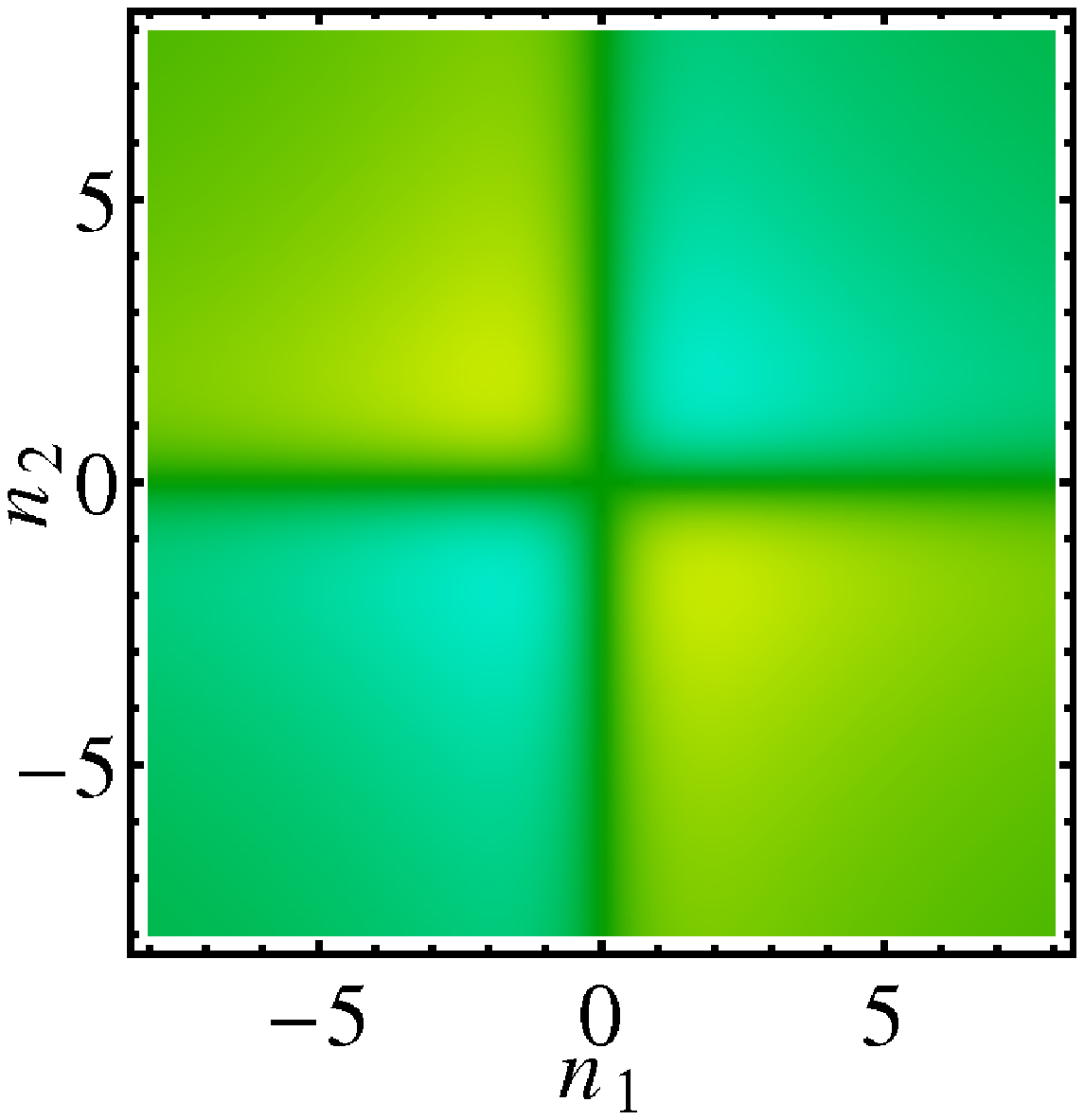,width=3.5cm} &
\hspace*{-0.5cm}\epsfig{file=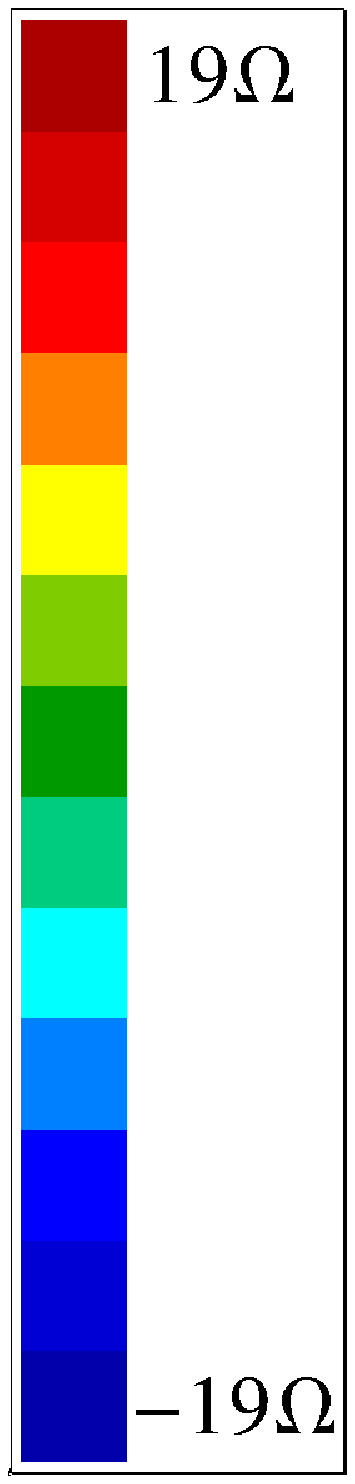,width=0.8cm}
\\
\epsfig{file=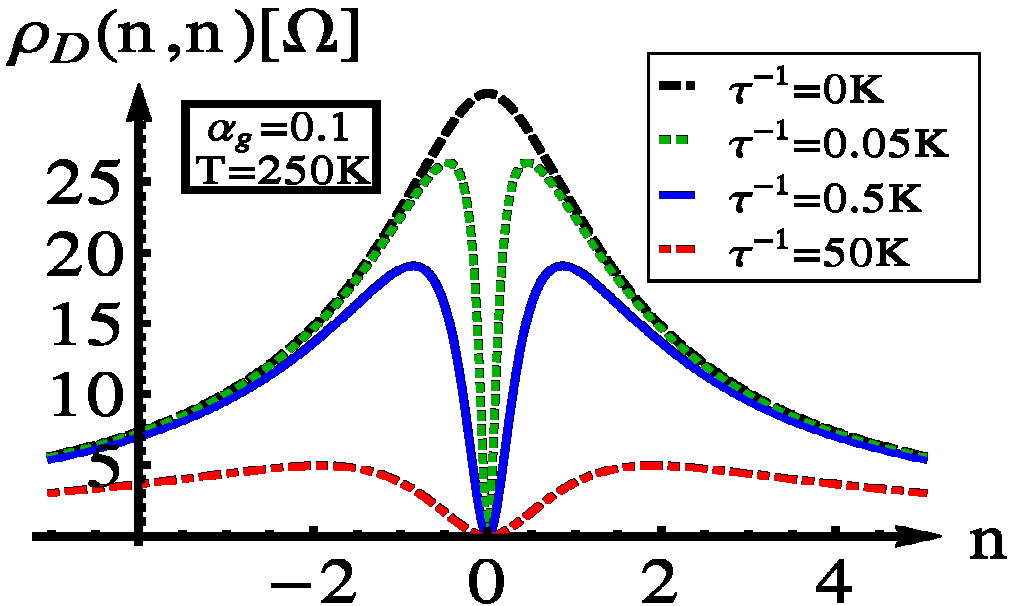,width=3.6cm} &
\epsfig{file=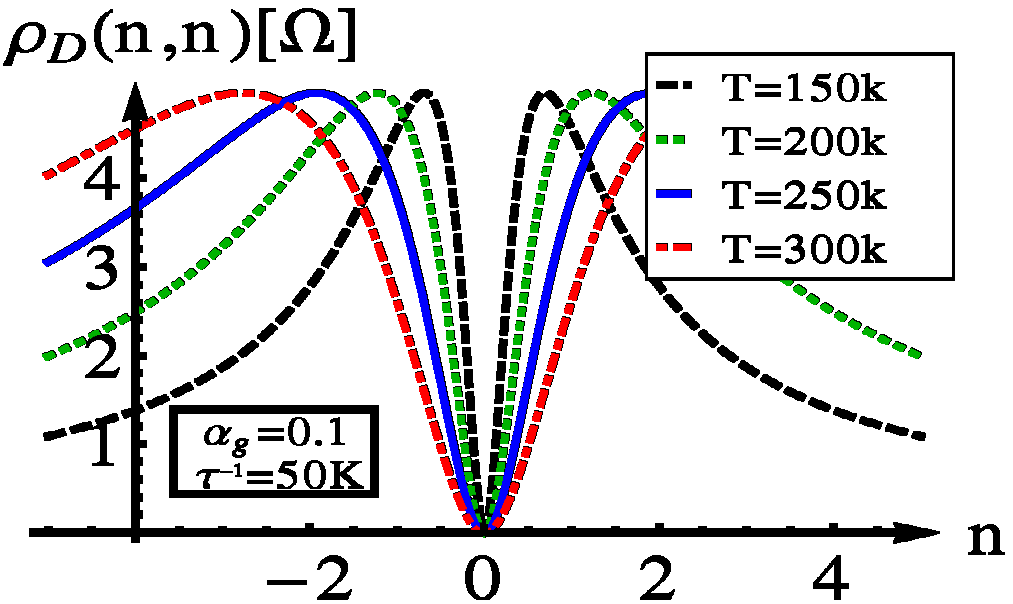,width=3.6cm} &
\end{array}$
\caption{(Color online) Drag coefficient in the ballistic regime as a
  function of carrier densities (in units of $10^{11}$ cm$^{-2}$) for
  $d=9$ nm. The left panels show $\rho_D$ at $T=250$ K, with the upper
  panel corresponding to ultra-clean graphene $\tau^{-1}=0.5$ K and
  the lower left panel showing the evolution of $\rho_D$ with
  increasing disorder from $\tau^{-1}=0$ to $\tau^{-1}=50$ K. The
  right panels show $\rho_D$ for $\tau^{-1}=50$ K. The four curves on
  the lower panel correspond to $T=150$ K, $200$ K, $250$ K, and $300$
  K.}
\label{fr1}
\end{figure}

The physics of Coulomb drag is well understood if both layers are
in the Fermi liquid state \cite{kor,fl2}. The electric field in the
passive layer is induced by exciting pairs of electron-like and
hole-like excitations in a state with finite total momentum. The
momentum is transferred from the current-carrying state in the active
layer by the inter-layer Coulomb interaction. The inter-layer momentum
transfer can be described by the effective relaxation rate
$\tau_D^{-1}$. The most basic qualitative features of the drag
measurement \cite{roj,kor,fl2} can already be inferred by estimating
$\tau_D^{-1}$ with the help of Fermi's golden rule, where it is
crucial to take into account the energy dependence of the density of
states (DoS) and/or diffusion coefficient $D$: indeed, the
current-carrying states can be characterized by non-zero total
momentum only in the case of electron-hole asymmetry.

The drag coefficient $\rho_D$ and momentum relaxation rate
$\tau_D^{-1}$ can be related using a simple Drude-like model.
Consider the phenomenological equations of motion, assuming for
simplicity that both layers are characterized by the same carrier
density $n$ and effective mass $m$
\begin{equation}
\label{emp}
\frac{d}{dt}
\begin{pmatrix}
\bs{j}_1 \cr
\bs{j}_2
\end{pmatrix}
= \frac{e^2n}{m}
\begin{pmatrix}
\bs{E}_1 \cr
\bs{E}_2
\end{pmatrix} - 
\frac{1}{\tau_D}
\begin{pmatrix}
1 & -1 \cr
-1 & 1
\end{pmatrix}
\begin{pmatrix}
\bs{j}_1 \cr
\bs{j}_2
\end{pmatrix}
- \frac{1}{\tau}
\begin{pmatrix}
\bs{j}_1 \cr
\bs{j}_2
\end{pmatrix}
,
\end{equation}
where $\bs{j}_{1(2)}$ is the average current density in the active (passive)
layer, $\bs{E}_{1(2)}$ is the electric field in the two layers, and
$\tau$ is the impurity scattering time. Noting that in the drag
measurement no net current is allowed to flow in the passive layer
$\bs{j}_2=0$, we arrive at the Drude-like formula
\begin{equation}
\label{dfd}
\rho_D = -\rho_{12} = \left( e^2 n \tau_D / m \right)^{-1}.
\end{equation}
Combining Eq.~(\ref{dfd}) with the Fermi's golden rule estimate for
$\tau_D^{-1}$ one can estimate the drag coefficient. More rigorous
calculations based on either the diagrammatic perturbation theory
\cite{kor} or the kinetic equation \cite{fl2} confirm the
``Fermi-liquid'' result
\begin{equation}
\label{fld}
\rho_D^{FL} =  (\hbar/e^2) A_{12} T^2/(\mu_1\mu_2) ,
\end{equation}
where $\mu_{1(2)}$ is the chemical potential of the active (passive)
layer and $A_{12}$ is determined by the matrix elements of the
inter-layer interaction (the precise form of $A_{12}$
as a function of the inter-layer spacing $d$ depends on whether
transport in the two layers is ballistic or diffusive \cite{kor}).

Even though the drag coefficient (\ref{fld}) is apparently independent
of the impurity scattering time $\tau$, transport properties of each
individual layer are usually \cite{roj,kor} assumed to be dominated by
disorder, $\tau\ll\tau_D$. In particular, solving Eq.~(\ref{emp}) for
the resistivity one finds the usual Drude formula. In contrast, the
behavior of clean double-layer systems, i.e. with $\tau\gg\tau_D$, is
less trivial. In this case, the last term in Eq.~(\ref{emp})
may be neglected leading to the non-zero result for the single-layer
resistivity
\begin{equation}
\label{rr}
\rho_{11} = - \rho_{12} = \left( e^2 n\tau_D/m \right)^{-1} = \rho_D.
\end{equation} 
Note, that the system is still characterized by the infinite
conductivity ($\hat\rho^{-1}=\infty$), as expected for disorder-free
conductors on the grounds of Galilean invariance.

The physical picture of the drag effect outlined so far is based on
the following assumptions: (i) each of the layers is assumed to be in
a Fermi-liquid state, which at the very least means $\mu_{1(2)}\gg T$;
(ii) electron-electron interaction does not contribute to the
transport scattering time; (iii) the inter-layer Coulomb interaction
is assumed to be weak enough, $\alpha = e^2/(\hbar v_F)\ll 1$, such
that $\rho_D$ is determined by the lowest-order perturbation theory
\cite{kor}.

Lifting one or more of the above assumptions leads to significant
changes in the drag effect \cite{ex2,ex3,ex4,tut,tu2,ge2}. In this
Letter we focus on the system of two parallel graphene sheets
\cite{tut,tu2,ge2,me1,dsa,csn,ds2,tud,us1,pet,per,sch,glz}, which offers a
great degree of control over the microscopic structure of the two
layers. Indeed, using hexagonal boron nitride as a substrate
\cite{ge2,ge1}, one can decrease disorder strength in the system and
reach the regime, where transport properties of the two layers are
dominated by electron-electron interaction,
$\tau\gg\tau_{ee}$. Moreover, the carrier density can be
electrostatically controlled allowing one to scan a wide range of
chemical potentials from the Fermi liquid regime to the Dirac point.

While inapplicable to massless fermions in graphene, the equations of
motion (\ref{emp}) provide an expectation of non-zero resistance in
the case of the ultra-clean system. Below, we use the quantum kinetic
equation (QKE) approach \cite{kas,kin} to derive hydrodynamic
equations \cite{ryz} that generalize Eq.~(\ref{emp}) for interacting Dirac
fermions in graphene. Solving these equations (or equivalently, the
QKE) we confirm that the system of two ultra-clean graphene sheets is
indeed characterized by a non-zero, but degenerate resistance matrix
whose elements satisfy Eq.~(\ref{rr}), with $\rho_D$ shown in
Fig.~\ref{fr1}.

\begin{figure}
\begin{center}
\epsfig{file=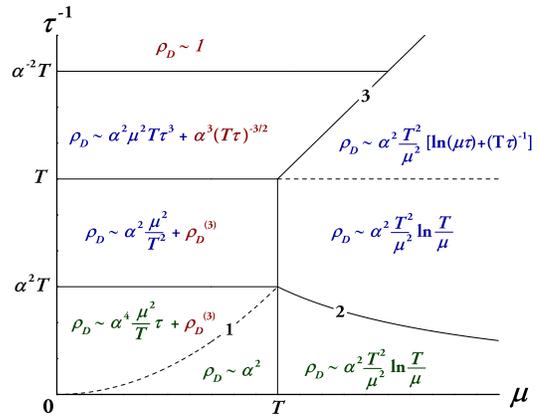,width=7cm}
\end{center}
\caption{(Color online) Drag coefficient in the case of identical
  layers in different parameter regimes for $\mu \ll {\rm
    min}(T/\alpha, v/d)$. The bottom row of results (for
  $\tau^{-1}\ll\alpha^2T$ and below the curve 2,
  $\tau^{-1}\ll\alpha^2T^2/\mu$) are obtained from the solution of the
  QKE (\ref{gr}). The curve 1 ($\tau^{-1}=\alpha^2\mu^2/T$) separates
  the two regimes in Eq.~(\ref{r1dp}). The middle row of results (for
  $\alpha^2T\ll\tau^{-1}\ll T$) corresponds to the region where the
  results and the applicability of the QKE overlap with those of the
  perturbation theory of Ref.~\onlinecite{us1} (for $\mu\gg T$ the
  results above and below the curve 2 contain different numerical
  factors). The third-order contribution $\rho_D^{(3)}={\cal
    O}(\alpha^3)$ resulting in small non-zero drag at $\mu=0$ is shown
  in red. The upper row of results ($\tau^{-1}\gg T$) corresponds to
  the diffusive regime [see Eqs.~(\ref{r3dp}) and (\ref{rkl})], where
  $\rho_D^{(3)}$ saturates for $\tau^{-1}\gg T/\alpha^2$).}
\label{pd}
\end{figure}

{\em Kinetic equation. ---} We now briefly outline the derivation of
the QKE for double-layer graphene structures and its solution in
the ballistic regime (see Supplemental Material \cite{sup}). Consider
an infinite sample in an infinitesimal, homogeneous electric field
$\bs{E}_1$ applied to the active layer. The response of the system to
the field can be described by the small non-equilibrium corrections
$h_{1(2)}$ to the Fermi distribution functions defined by
\begin{equation}
n_i(\epsilon,\hat{\mathbf{v}})=n_F^{(i)}(\epsilon) + 
T \frac{\partial n_F^{(i)}(\epsilon)}{\partial \epsilon} h_i(\epsilon,\hat{\mathbf{v}}),
\label{h}
\end{equation}
where the eigenstates of the Dirac Hamiltonian $H = v \bs{\sigma}
\bs{p}$ are labeled \cite{sup} by their energy $\epsilon$ and the
velocity unit vector $\hat{\mathbf{v}}$; the momentum of the particle
is $\bs{p} = \epsilon \hat{\mathbf{v}}/v$.

Small corrections $h_{1(2)}$ can be found by linearizing the QKE
\cite{fn2}
\begin{eqnarray}
&&
\frac{\partial h_1}{\partial t}+\frac{e\bs{E}_1\bs{v}}{T}=
-\frac{h_1}{\tau} + I_{11}\{h_1\} + I_{12}\{h_1,h_2\},
\nonumber\\
&&
\nonumber\\
&&
\frac{\partial h_2}{\partial t}=-\frac{h_2}{\tau} 
+ I_{22}\{h_2\} + I_{21}\{h_2,h_1\},
\label{ke1}
\end{eqnarray} 
where the linearized pair-collision integrals are given by
\begin{eqnarray}
\label{i}
&&
I_{ij}= - \int d2\ d3\ d4\ 
W^{ij}(h_{i,1}-h_{i,2}+h_{j,3}-h_{j,4}),
\nonumber\\
&&
\nonumber\\
&&
W^{ij}=
\delta(\bs{p}_1-\bs{p}_2+\bs{p}_{3}-\bs{p}_{4})
 \ 
\delta(\epsilon_1-\epsilon_2+\epsilon_{3}-\epsilon_{4})
\nonumber\\
&&
\nonumber\\
&&
\quad
\times 
\frac{ \cosh\frac{\epsilon_1-\mu_i}{2T}}{2 \cosh\frac{\epsilon_2-\mu_i}{2T}
\cosh\frac{\epsilon_3-\mu_j}{2T}\cosh\frac{\epsilon_4-\mu_j}{2T}}
K^{ij}_{1,2;3,4},
\end{eqnarray}
and we have used short-hand notations
$h_{i,a}=h(\epsilon_a,\hat{\mathbf{v}}_a)$,
$da=\nu(\epsilon_a) d\hat{\mathbf{v}}_ad\epsilon_a$, with $a=1,2,3,4$. The kernel
\begin{equation}
K^{ij}_{1,2;3,4}=|U^{ij}(\bs{p}_1-\bs{p}_2)|^2
\frac{1+\hat{\mathbf{v}}_1\hat{\mathbf{v}}_2}{2}
\frac{1+\hat{\mathbf{v}}_3\hat{\mathbf{v}}_4}{2},
\label{k}
\end{equation}
contains the interaction matrix element describing the two-particle
scattering $1\to 2 $ and $3\to 4$ and the corresponding Dirac
factors. Here we take into account only the Hartree interaction term:
there is no exchange interaction between the layers, whereas within
the layers the Hartree term dominates in the large-$N$ limit ($N$ is
the number of electron flavors; physically, $N=4$ due to spin and
valley degeneracy).

The peculiarity of the inelastic scattering in the Dirac spectrum is
two-fold. First, since the velocity $\bs{v}=v^2\bs{p}/\epsilon$ is
independent of the absolute value of the momentum, total momentum
conservation does not prevent velocity (or current) relaxation. As
a result, the intralayer collision integral $I_{ij}$ yields a non-zero
transport relaxation rate due to electron-electron scattering.

Second, the scattering of particles with almost collinear momenta is
enhanced since the momentum and energy conservation laws coincide for
collinear scattering. This restricts the kinematics \cite{kas,kin,po2} of
the Dirac fermions leading to the singularity in the collision
integral. This singularity leads to the fast thermalization of
particles within a given direction, which justifies the Ansatz:
\begin{equation}
\label{Ansatz}
h_i(\epsilon,\hat{\bf{v}})=\left(\chi_v^{(i)}+\chi_p^{(i)} \;
\epsilon/T\right) e\bs{E}\bs{v}/T^2.
\end{equation}
The Ansatz~(\ref{Ansatz}) retains the only two modes for which the
collision integral $I_{ij}$ is not singular: the ``momentum mode''
$\chi_p^{(i)}$, which nullifies the collision integral due to momentum
conservation, and the ``velocity mode'' $\chi_v^{(i)}$, which
nullifies $I_{ij}$ in the case of collinear scattering. The same
kinematic restrictions lead to fast uni-directional thermalization
between the layers. This allows us to set $\chi_p^{(1)}=\chi_p^{(2)}$,
and hence reduce the QKE for the double-layer setup to a
$3\times3$ matrix equation.

Consider for simplicity the case of identical layers (for the more
general case of $\mu_1\ne \mu_2$ see Supplemental Material
\cite{sup}).  Integrating the reduced QKE over the
energies, we arrive at the set of steady-state hydrodynamic equations
in terms of the particle currents
\begin{equation}
\bs{J}_i = - N T \int d\epsilon 
\nu(\epsilon)  
\frac{\partial n_F^{(i)}}{\partial \epsilon} 
\int d\hat{\mathbf{v}} \bs{v} h_i(\epsilon,\hat{\bf{v}}),
\end{equation}
and the total momentum $\bs{P}=e\epsilon_0C_1^2(\bs{E}_1+\bs{E}_2)\tau$:
\begin{equation}
\label{heq}
e\epsilon_0 
\begin{pmatrix}
\bs{E}_1 \cr 
\bs{E}_2 
\end{pmatrix}
=
\left[
\frac{1}{\tau} + \widehat{\cal I}_{ee} - \widehat{\cal I}_D\right]
\begin{pmatrix}
\bs{J}_1 \cr 
\bs{J}_2 
\end{pmatrix}
+\left[\frac{1}{\tau_{D}}-\frac{1}{\tau_{ee}}\right] 
\begin{pmatrix}
\bs{P} \cr \bs{P}
\end{pmatrix},
\end{equation}
where $\widehat{\cal
  I}_{ee(D)}=[(\hat\sigma_0+\hat\sigma_1)C_1^2+2\hat\sigma_{0(1)}C_2]/\tau_{ee(D)}$,
the intra- and inter-layer electron-electron transport scattering
rates are (${\cal W}^{ij} = W^{ij} \nu_1/\cosh^2[(\epsilon_1-\mu_i)/(2T)]$)
\begin{eqnarray}
&&
\frac{1}{\tau_{D}}=\frac{1}{4T\epsilon_0C_2} 
\int \prod_{a=1}^4 da {\cal W}^{12}
\left(\bs{v}_1-\bs{v}_2\right)\left(\bs{v}_4-\bs{v}_3\right),
\nonumber\\
&&
\nonumber\\
&&
\frac{1}{\tau_{ee}}=\frac{1}{8T\epsilon_0C_2} 
\int \prod_{a=1}^4 da
\left[ {\cal W}^{11}
\left(\bs{v}_1-\bs{v}_2+\bs{v}_3-\bs{v}_4\right)^2 \right.
\nonumber\\
&&
\nonumber\\
&&
\quad\quad\quad\quad\quad\quad\quad\quad\quad\quad\quad
\left.
+
2 {\cal W}^{12}\left(\bs{v}_1-\bs{v}_2\right)^2 \right],
\end{eqnarray}
and $\sigma_k$ are the Pauli matrices in ``layer space''. The
coefficients $C_{1(2)}$ represent the average energy and energy
variation, while $\epsilon_0= 2T {\cal J}\{1\}/N$ is a typical energy:
\begin{eqnarray*}
&&
C_1 = \frac{\left\langle \epsilon \right\rangle_\epsilon}{T}\sim\frac{\mu}{T}, \quad
C_2 = \frac{\left\langle\epsilon^2\right\rangle_\epsilon-
\left\langle \epsilon \right\rangle^2_\epsilon}{T^2} \sim const,
\nonumber\\
&&
\nonumber\\
&&
{\cal J}\{\dots\}=
-\frac{v^2}{T}\int d\epsilon 
\nu(\epsilon)  
\frac{\partial n_F}{\partial \epsilon} \dots , \quad
\left\langle\dots\right\rangle_\epsilon = \frac{{\cal J}\{\dots\}}{{\cal J}\{1\}}.
\end{eqnarray*}
The hydrodynamic equations (\ref{heq}) generalize the equations of
motion (\ref{emp}) to the case of Dirac fermions in graphene. The
kinematic peculiarity of Dirac fermions manifests itself in the
appearance of the total momentum, which entangles the electric fields
in the two layers.

Solving the hydrodynamic equations (\ref{heq}) we find 
\begin{equation}
\label{gr}
\rho_D =\frac{\hbar}{e^2}\frac{C_2}{\epsilon_0}
\frac{(\tau\tau_D)^{-1}+C_1^2\left[\tau^{-2}_{ee}-\tau^{-2}_{D}\right]}
{\tau^{-1}+C_1^2\left[\tau_{ee}^{-1}-\tau_D^{-1}\right]}.
\end{equation}
For a clean system, the resistivity matrix is degenerate and the
drag coefficient is given by
\begin{equation}
\rho_D (\tau\to\infty)=(\hbar/e^2)
(C_2/\epsilon_0) \left(\tau_D^{-1}+\tau_{ee}^{-1}\right),
\end{equation}
which remains non-zero $\rho_D\sim(\hbar/e^2)\alpha^2$ even at the
Dirac point $\mu=0$, where it is determined by
$\tau_{ee}^{-1}\sim\alpha^2T$ (to the second order in the inter-layer
interaction $\tau_D^{-1}(\mu=0)=0$, while the third-order contribution
$\tau_D^{-1}(\mu=0)\sim\alpha^3T$ is subleading; the latter is
expected to dominate the effect for sufficiently strong disorder, see
below).

Equation~(\ref{gr}) gives the general expression for the drag
coefficient in the ballistic regime based on the solution of the
QKE (\ref{ke1}). For arbitrary parameter values this
expression is to be evaluated numerically (see Fig.~\ref{fr1} for the
numerical results). Analytical expressions can be obtained for various
limiting cases (summarized in Fig.~\ref{pd}). Below, we discuss the
asymptotic behavior of $\rho_D$ for the case of two inequivalent
layers \cite{sup} focusing on the experimentally relevant case
\cite{tut,tu2,ge2} $Td/v<1$ and analyzing the evolution of $\rho_D$
with increasing disorder strength.

{\em Ballistic regime. ---} For weak disorder $\alpha^2T\tau\gg 1$ (or
$\tau^{-1}\ll\tau_{ee}^{-1}$) and neglecting the third-order
contribution to $\tau_D^{-1}$, we find for $\rho_D$ near the Dirac
point 
\begin{equation}
\label{r1dp}
\rho_D (\mu_i\ll T) \approx 2.87
\frac{h}{e^2} \; \alpha^2 
\frac{\mu_1\mu_2}{\mu_1^2+\mu_2^2+0.49T/(\alpha^2\tau)},
\end{equation}
where $\tau_D^{-1}\sim\alpha^2\mu_1\mu_2/T$,
$\tau_{ee}^{-1}\sim\alpha^2T$, $\epsilon_0\sim T$, and $C_2\sim 1$.
The value of $\rho_D$ precisely at the Dirac point depends on the
experimental set-up. For clean samples, if one of the chemical
potentials remains non-zero, while the other is scanned through the
Dirac point \cite{tut}, then $\rho_D(\mu_1=0, \mu_2\ne 0)=0$, similar
to Ref.~\onlinecite{tu2}. On the contrary, if both chemical potentials
are driven through the Dirac point simultaneously \cite{ge2}, then
Eq.~(\ref{r1dp}) predicts a non-vanishing value of
$\rho_D(\mu_1=\pm\mu_2=0)\ne 0$, see Fig.~\ref{fr1}.

For intermediate disorder strength $\alpha^2 T \ll \tau^{-1} \ll T$
the applicability region of the QKE overlaps with
that of the conventional perturbation theory developed in
Ref.~\onlinecite{us1} and we recover perturbative results, see
Fig.~\ref{pd}.

For even stronger disorder (or at low temperatures) $T\tau\ll 1$ the
electron motion becomes diffusive. In this case the kinematic
restrictions are relaxed and the Ansatz (\ref{Ansatz}) is no longer
justified.  However, in this regime, the perturbative approach is
applicable and allows for a standard description of the diffusive
transport.

\begin{figure}[h]
\begin{center}
\epsfig{file=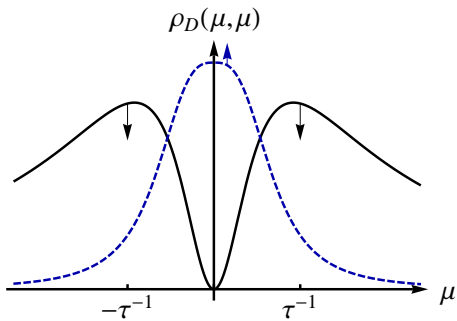,width=6cm}
\end{center}
\caption{(Color online) Schematic view of the drag coefficient at low
  temperatures: the second-order contribution $\rho_D^{(2)}$ (solid
  line) and the third-order contribution $\rho_D^{(3)}$ (blue dashed
  line). The arrows indicate the tendency of the two terms with the
  decrease of temperature $T\to 0$.}
\label{rdif}
\end{figure}

{\em Diffusive regime. ---} 
The lowest-order perturbative calculation \cite{kor} amounts to
evaluation of the Aslamasov-Larkin-type diagram for the drag
conductivity given by
\begin{equation}
\label{sd}
\sigma^{\alpha\beta}_D = \frac{1}{16\pi T}
\sum_{\bs{q}}
\int \frac{d\omega}{\sinh^2\frac{\omega}{2T}}
\Gamma_1^\beta(\omega, \bs{q}) 
\Gamma_2^\alpha(\omega, \bs{q}) 
| {\cal D}^R_{12} |^2,
\end{equation}
where ${\cal D}^R_{12}$ is the retarded propagator of the inter-layer
interaction and $\Gamma_a^\alpha(\omega, \bs{q})$ is the non-linear
susceptibility [in fact, all previous studies of the Coulomb drag in
  graphene \cite{me1,dsa,csn,ds2,us1,tud,pet,per,sch} focused on
  Eq.~(\ref{sd})]. In the diffusive regime, $\Gamma_a^\alpha(\omega, \bs{q})$
can be found using the Ohm's law and the continuity
equation \cite{nas} $\bs{\Gamma} = e \bs{q} (\partial\sigma/\partial
n) {\rm Im} \Pi^R$. All microscopic details are now encoded in the
diffusion coefficient and the density dependence of the single-layer
conductivity $\sigma$.  Close to the Dirac point $\mu\ll T\ll
\tau^{-1}$ the derivative $\partial\sigma/\partial n \sim n v^2\tau^2$
(independently of the precise nature of impurities). After this the
evaluation of Eq.~(\ref{sd}) is rather standard (except that, in
contrast to Ref.~\onlinecite{kor}, the Thomas-Fermi screening length
is much longer than the inter-layer spacing $\varkappa d\ll 1$) and
yields
\begin{equation}
\label{r3dp}
\rho_D^{(2)}\left(\mu_i\ll T \ll\tau^{-1}\right) \sim
(\hbar/e^2) \alpha^2 \mu_1\mu_2 T \tau^3.
\end{equation}
This result vanishes at the Dirac point as a consequence of the
electron-hole symmetry.

The importance of the electron-hole asymmetry for the Coulomb drag
follows from Eq.~(\ref{sd}): the non-linear susceptibility can be
thought of as a measure of the asymmetry. However, Eq.~(\ref{sd}) is
only the lowest-order contribution to $\sigma_D$. Under standard
assumptions of the Fermi-liquid behavior in the two layers ($\mu\gg
v/d \gg T$, $\mu\tau\gg 1$), this contribution indeed dominates the
observable effect. On the contrary, in the vicinity of the Dirac point
in graphene, the next-order contribution $\rho_D^{(3)}$ \cite{lak}
becomes important since it is insensitive to the electron-hole
symmetry and thus does not vanish at the Dirac point.

The explicit results of Ref.~\onlinecite{lak} were obtained in the
usual limit $\varkappa d\gg 1$. Extending these calculations to the
opposite case $\varkappa d\ll 1$ we find close to the Dirac
point 
\begin{equation}
\label{rkl}
\rho_D^{(3)}\left(\mu_i\ll T\ll\tau^{-1}\ll\alpha^{-2}T\right) 
\sim (\hbar/e^2)\alpha^3(T\tau)^{-3/2},
\end{equation}
and $\rho_D^{(3)}\sim \hbar/e^2$ for $\tau^{-1}\gg\alpha^{-2}T$. Away
from the Dirac point this contribution decays as a function of the
chemical potential $\rho_D^{(3)}(\mu\tau\gg{\rm
  max}[1,\alpha^{-1}(T\tau)^{1/2}])\sim(\hbar/e^2)(\mu\tau)^{-3}$ and
rapidly becomes subleading. As a result, $\rho_D^{(3)}$ is only
detectable at low $T$ and $\mu$, see Fig.~\ref{rdif}.

While estimating $\rho_D^{(3)}$ at the Dirac point, we assume the
single-layer conductivity $\sigma\sim e^2/h$ discarding localization
effects. Indeed, experiments on high-quality samples show
$T$-independent $\sigma$ down to $T=30$ mK \cite{kim}, that can be
explained by the specific character of disorder in graphene \cite{ogm}.

{\em Summary. ---} We have studied Coulomb drag in double-layer
graphene structures. By using the QKE formalism we have shown that for
weak disorder (or high $T$; ballistic regime) $\rho_D$ near the Dirac
point is given by Eq.~(\ref{r1dp}), see also Fig.~\ref{fr1}, which is
consistent with Ref.~\onlinecite{tu2}. For $\alpha^2 T\tau\ll 1$, the
solution of the QKE agrees with the perturbative calculation of
Ref.~\onlinecite{us1}. For even stronger disorder (or lower $T$;
diffusive regime) subleading third-order contribution dominates the
effect in qualitative agreement with the experimentally observed peak
at the Dirac point at low temperatures \cite{ge2}.  A possible
alternative origin of the low-$T$ peak at the Dirac point is a
collective state of the double-layer system, either due to strong
interaction \cite{vin} (not explored here) or correlated disorder
\cite{gor,lev} (discussed in Supplemental Material \cite{sup}).

{\em Acknowledgements. ---} We thank A.K. Geim, K.S. Novoselov, and
L. Ponomarenko for communicating their experimental results prior to
publication. We are grateful to J. Schmalian and L. Fritz for
stimulating discussions. This research was supported by the Center for
Functional Nanostructures and SPP 1459 ``Graphene'' of the Deutsche
Forschungsgemeinschaft (DFG), and by BMBF. M.T. is grateful to KIT for
hospitality. After this work was completed, we became aware of a
closely related studies by J. Lux and L. Fritz \cite{lux}. 


\begin{widetext}

\appendix

\section{Coulomb drag in graphene near the Dirac point: Supplemental Material}

\section{1. Kinetic equation approach}
In this section we provide details of derivation of the resistivity
tensor in double layer graphene from the kinetic equation approach.

\subsection{A. Kinetic equation}
Eigenstates of the massless Dirac Hamiltonian $H = v \bm{\sigma} \mathbf{p}$
are characterized by the values of momentum $\mathbf{p}$ and the discrete
variable $\chi = \pm 1$ indexing conduction and valence bands. In this
representation, energy and velocity are $\epsilon = \alpha v |\mathbf{p}|$ and
$\mathbf{v} = \chi v \mathbf{p}/p$.
It is, however, more convenient to label the eigenstates by their energy $\epsilon$
and the unit velocity vector $\hat{\mathbf{v}}$. The momentum of the particle is
then $\mathbf{p} = \epsilon \hat{\mathbf{v}}/v$ and the normalization of the
states reads
\begin{equation}
 \int \frac{|\epsilon|\, d\epsilon\, d\hat{\mathbf{v}}}{(2\pi v)^2}\;
 | \epsilon, \hat{\mathbf{v}} \rangle \langle \epsilon, \hat{\mathbf{v}} |
  = 1.
\end{equation}

We consider an infinite double-layer graphene sample (layers $1$ and $2$)
in a homogeneous electric field $\mathbf{E}_1$ applied to the active layer $1$.
Assuming weak electric field, we start with the linearized kinetic equation:
\begin{eqnarray}
&&\frac{\partial h_1}{\partial t}+\frac{e\mathbf{E}_1\mathbf{v}}{T}=
-\frac{h_1}{\tau} + I_{11}\{h_1\} + I_{12}\{h_1,h_2\},
\nonumber\\
&&
\nonumber\\
&&
\frac{\partial h_2}{\partial t}=-\frac{h_2}{\tau} + I_{22}\{h_2\} + I_{21}\{h_2,h_1\}.
 \label{kineq1}
 \end{eqnarray}
Here the nonequilibrium correction $h_i$ to the Fermi distribution function
is defined by
 \begin{equation}
 n_i(\epsilon,\hat{\mathbf{v}})=n_F(\epsilon) + T \frac{\partial n_F(\epsilon)}{\partial \epsilon} h_i(\epsilon,\hat{\mathbf{v}}),
 \label{eq3}
 \end{equation}
 and $I_{ij}$ is the linearized pair-collision integral:
\begin{equation}
I_{ij} = - N \int d\epsilon_2d\epsilon_3d\epsilon_4  
\int d\hat{\mathbf{v}}_2 d\hat{\mathbf{v}}_3 d\hat{\mathbf{v}}_4
\nu(\epsilon_2)\nu(\epsilon_3)\nu(\epsilon_4)
 W^{ij}(1,3;2,4)\ \left[(h_{i}(\epsilon_1,\mathbf{v}_1)-h_{i}(\epsilon_2,\mathbf{v}_2)
 +h_{j}(\epsilon_3,\mathbf{v}_3)-h_{j}(\epsilon_4,\mathbf{v}_4)\right],
 \label{eq2}
 \end{equation}
\begin{equation}
W^{ij}(1,2;3,4)=\delta(\mathbf{p}_1-\mathbf{p}_2+\mathbf{p}_{3}-\mathbf{p}_{4})
 \ \delta(\epsilon_1-\epsilon_2+\epsilon_{3}-\epsilon_{4})\  
\frac{\cosh\frac{\epsilon_1-\mu_i}{2T}}
{2 \cosh\frac{\epsilon_2-\mu_i}{2T}\cosh\frac{\epsilon_3-\mu_j}{2T}\cosh\frac{\epsilon_4-\mu_j}{2T}}
 K^{ij}(1,2;3,4).
 \label{eq4}
\end{equation}
Here $\nu(\epsilon)$ is the density of states for one of $N$ flavors
(per spin and per valley in graphene, where $N=4$). We assume formally
the large $N$ limit and neglect the intralayer exchange interaction.
We further assume that the scattering does not mix flavors (i.e., we
neglect the intervalley scattering due to Coulomb interaction): states
$1(3)$ and $2(4)$ belong to the same flavor, which gives the overall
factor $N$.  The kernel
 \begin{equation}
 K^{ij}(1,2;3,4)=|M^{ij}|^2
 \frac{1+\hat{\mathbf{v}}_1\hat{\mathbf{v}}_2}{2}\frac{1+\hat{\mathbf{v}}_3\hat{\mathbf{v}}_4}{2}
\label{Kernel}
 \end{equation}
contains the interaction matrix element $M_{ij}$ describing the collision of two particles
$1\to 2 $ and $3\to 4$ and the corresponding Dirac factors.
Within the Golden-rule approximation, this matrix element is given by the Fourier component of the
interaction potential:
\begin{equation}
M_{ij}^{(1)}=U_{ij}^{(0)}(\mathbf{p}_1-\mathbf{p}_2),
\end{equation}
where
\begin{equation}
\hat{U}^{(0)}(q)= V_0(q)
\begin{pmatrix}
1 & e^{-q d}\\
e^{-q d} & 1
\end{pmatrix},
\end{equation}
with
\begin{equation}
V_0(q)=\frac{2\pi e^2}{q}.
\end{equation}
Further, one can generalize the collision integral to the case of the RPA-screened
interaction. Then
\begin{equation}
M_{ij}=U_{ij}^{\text{RPA}}(\mathbf{p}_1-\mathbf{p}_2, vp_1-vp_2),
\end{equation}
where
\begin{eqnarray}
\hat{U}^{\text{RPA}}(\mathbf{q}, \omega)
&=&\frac{V_0(q)}{\left[1+V_0(q)\Pi_1(q,\omega)\right]
\left[1+V_0(q)\Pi_2(q,\omega)\right]-e^{-2qd}V_0^2(q)\Pi_1(q,\omega)\Pi_2(q,\omega)}
\nonumber\\
&&
\nonumber\\
&\times&
\begin{pmatrix}
1+V_0(q)\Pi_2(q,\omega)\left(1-e^{-2qd}\right) & e^{-q d} \cr
e^{-q d} & 1+V_0(q)\Pi_1(q,\omega)\left(1-e^{-2qd}\right)
\end{pmatrix},
\label{RPA}
\end{eqnarray}
where $\Pi_i(q,\omega)$ is the polarization operator in layer $i$.

We will focus on the experimentally relevant case of closely located layers,
$Td/v\ll 1$. Furthermore, here we will restrict our consideration to the case of relatively
low concentrations, such that $\mu d/v\ll 1$ (the situation with large interlayer distance
will be considered in detail elsewhere). Under these conditions we can set $d=0$ in the
interaction matrix elements so that the intralayer and interlayer interactions are just the same.
We further assume that the interaction coupling constant is small 
\begin{equation}
\alpha=\frac{e^2}{v}\ll 1.
\end{equation}

For simplicity, we treat impurity scattering within the relaxation time
approximation with an energy-independent transport time $\tau$.
Generalization to the more realistic case of Coulomb impurities with
an energy-dependent transport time will be discussed elsewhere. 

\subsection{B. Collinear-scattering singularity}

The momentum and energy conservation establishes severe kinematic
restrictions on the scattering in systems with linear spectrum \cite{kas}.
This can be easily seen, when one rewrites the product of delta-functions in Eq. (\ref{eq4}) as integrals over 
 $\mathbf{q}=\mathbf{p}_1-\mathbf{p}_2$, $\omega=\epsilon_1-\epsilon_2$,
 and two angles $\phi_{1(3)}$ between $\mathbf{q}$ and $\mathbf{p}_{1(3)}$, respectively:
 \begin{eqnarray}
 \delta(\mathbf{v}_1\epsilon_1-\mathbf{v}_2\epsilon_2+\mathbf{v}_{3}\epsilon_{3}-\mathbf{v}_{4}\epsilon_{4})
 &=&\int \frac{d^2 q}{(2\pi)^2} \delta(\mathbf{v}_1\epsilon_1-\mathbf{v}_2\epsilon_2-\mathbf{q})
 \delta(\mathbf{v}_{3}\epsilon_{3}-\mathbf{v}_{4}\epsilon_{4}+\mathbf{q}),\nonumber \\
 \delta(\epsilon_1-\epsilon_2+\epsilon_{3}-\epsilon_{4})&=&\int_\infty^\infty
 d\omega \delta(\epsilon_1-\epsilon_2-\omega)\delta(\epsilon_{3}-\epsilon_{4}+\omega),\nonumber \\
 \label{qw}
 \end{eqnarray}
 This allows one to integrate out 
 $\mathbf{p}_2$ and $\mathbf{p}_4$ ($\mathbf{v}_{2,4}$ and $\epsilon_{2,4}$) in the collision 
 integral, leading to the product
 \begin{equation}
 \delta(\epsilon_1-\omega-\sqrt{\epsilon_1^2+q^2v^2-2\epsilon_1 qv \cos\phi_1})
 \delta(\epsilon_3+\omega-\sqrt{\epsilon_3^2+q^2v^2+2\epsilon_3 qv \cos\phi_3}),
 \end{equation}
 which sets 
 \begin{eqnarray}
 \epsilon_1-\omega=v|\mathbf{p}_1-\mathbf{q}| \ &\Rightarrow & \ \cos\phi_1=\frac{q^2v^2-\omega^2+2\epsilon_1\omega}{2\epsilon_1 q v},
 \\
 \epsilon_3+\omega=|\mathbf{p}_3+\mathbf{q}| \ &\Rightarrow & \ \cos\phi_3=\frac{\omega^2-q^2v^2+2\epsilon_3\omega}{2\epsilon_3 q v}.
  \label{eq11}
 \end{eqnarray} 
 When calculating the scattering rates using the collision integral (\ref{eq2}),
 the angular integration over $\phi_1$ and $\phi_3$ removes the delta-functions, producing
 the factor
 \begin{eqnarray}
 \frac{\epsilon_1-\omega}{\epsilon_1 q v |\sin\phi_1|} \frac{\epsilon_3+\omega}{\epsilon_3 q v |\sin\phi_3|}
 &=&\frac{4(\epsilon_1-\omega)(\epsilon_3+\omega)}
 {(q^2v^2-\omega^2)\sqrt{[(2\epsilon_1-\omega)^2-q^2v^2][(2\epsilon_3+\omega)^2-q^2v^2]}}.
  \label{eq16}
 \end{eqnarray}
 In combination with the Dirac factors from Eq. (\ref{Kernel}),
 \begin{eqnarray}
 (1+\mathbf{v}_1\mathbf{v}_2)(1+\mathbf{v}_3\mathbf{v}_4)
 =\frac{(2\epsilon_1-\omega)^2-q^2v^2}{2 \epsilon_1(\epsilon_1-\omega)}\
 \frac{(2\epsilon_3+\omega)^2-q^2v^2}{2 \epsilon_3(\epsilon_3+\omega)},
 \label{eq14}
 \end{eqnarray}
 this yields
 \begin{equation}
 \frac{1}{q^2v^2-\omega^2}\frac{\sqrt{(2\epsilon_1-\omega)^2-q^2v^2}}{\epsilon_1}\
 \frac{\sqrt{(2\epsilon_3+\omega)^2-q^2v^2}}{\epsilon_3}.
 \end{equation}
 Therefore, further integration over $q$ or $\omega$ generically
 produces a logarithmic divergence at $\omega=\pm qv$, which stems
 from the collinear scattering $\phi_3=\phi_1=0$ or $\pi$, see
 Eq. (\ref{eq11}) at the light cone. This divergence reflects the fact
 that for a linear spectrum the momentum and energy conservation laws
 coincide in the ``one-dimensional'' collinear case.  Note that this
 enhancement of the collinear scattering is not restricted to the case
 of undoped graphene.
 
 In order to regularize this divergence, one has to go beyond the
 Golden-rule level and take into account the screening of the
 interaction (which in the clean case is perfect exactly on the light
 cone) and renormalization of the spectrum due to interaction (leading
 to nonlinear corrections).  These mechanisms lead to the appearance
 of a large factor $|\ln(\alpha)|\gg 1$ in generic relaxation rates in
 graphene. In disordered graphene, this singularity is also cut off by
 disorder-induced broadening of the momentum-conservation
 delta-function.

\subsection{C. Ansatz}

The singularity in the collinear scattering in graphene leads to the
fast thermalization of particles within given direction within each of
the layers.  Clearly, in an external electric field $\mathbf{E}$, the
linearized nonequilibrium correction to the distribution function is
proportional to the driving term $\mathbf{E v}$.  Therefore, the
nonequilibrium correction
$$h_i(\epsilon,\mathbf{v})=\chi(\epsilon) \frac{e\mathbf{E v}}{T^2}$$
is characterized by some function of energy $\chi(\epsilon)$. One can
formally expand this function in $\epsilon/T$:
$$
\chi(\epsilon)=\sum_{n=0}^\infty\chi_n (\epsilon/T)^n.
$$
The action of the collision integral (which contains the combination $h_1-h_2+h_3-h_4$)
on this function generates the equilibration rate
in all terms except for $n=0,1$. Indeed, for $n=0$, the combination 
$$h^{(0)}_1-h^{(0)}_2+h^{(0)}_3-h^{(0)}_4 \propto
\chi_0(\mathbf{v}_1-\mathbf{v}_2+\mathbf{v}_3-\mathbf{v}_4)$$ contains
the differences of velocities.  This cancels the kinematics-induced
divergency $\propto
|\mathbf{v}_1-\mathbf{v}_2+\mathbf{v}_3-\mathbf{v}_4|^{-1}$.  For
$n=1$ the combination
\begin{eqnarray}
h^{(1)}_1-h^{(1)}_2+h^{(1)}_3-h^{(1)}_4  \propto \chi_1(\epsilon_1\mathbf{v}_1-\epsilon_2\mathbf{v}_2+\epsilon_3\mathbf{v}_3-\epsilon_4\mathbf{v}_4)
= \chi_1(\mathbf{p}_1-\mathbf{p}_2+\mathbf{p}_3-\mathbf{p}_4)
\end{eqnarray}
contains the change of the total momentum of two colliding particles,
which is exactly the argument of the momentum-conservation delta-function.
All other contributions with $n>1$ produce a relaxation rate enhanced by the
collinear scattering. The corresponding values of $\chi_n$ are therefore strongly suppressed
compared to $\chi_0$ and $\chi_1$, which justifies the following Ansatz \cite{kas}:
\begin{equation}\label{EqAnsatz}
 h_i(\epsilon,\mathbf{v})=\left(\chi_\mu^{i}+\chi_T^{i}\frac{\epsilon-\mu}{T}\right) \frac{e\mathbf{E}\mathbf{v}}{T^2}
 = \left(\chi_0^{i}+\chi_1^{i}\frac{\epsilon}{T}\right) \frac{e\mathbf{E}\mathbf{v}}{T^2}
\equiv \left(\chi_v^{i}+\chi_p^{i}\frac{\epsilon}{T}\right) \frac{e\mathbf{E}\mathbf{v}}{T^2}.
\end{equation}
This correction to the distribution function contains only the two
modes (proportional to velocity and momentum and characterized for
each layer by the two constants $\chi_v=\chi_0$ and $\chi_p=\chi_1$, respectively),
that nullify the collision integral in the case of collinear
scattering. The notation $\chi_\mu$ and $\chi_T$ is chosen to
emphasize that, after the linearization with respect to $\mathbf{E}$,
these quantities reflect the angular-dependent corrections to the
chemical potential and temperature, respectively, in the
direction-equilibrated distribution function:
\begin{eqnarray}
n(\epsilon,\hat{\mathbf{v}})&=&\frac{1}{1+\exp\left[\dfrac{\epsilon-\mu(\hat{\mathbf{v}})}{2 T(\hat{\mathbf{v}})}\right]}
\simeq n_F(\epsilon)-\frac{\partial n_F(\epsilon)}{\partial \epsilon}
\left[\frac{\delta\mu(\hat{\mathbf{v}})}{2T}+(\epsilon-\mu)\frac{\delta T(\hat{\mathbf{v}})}{2T^2}\right]
\\
&=&n_F(\epsilon)-\frac{1}{2T}\frac{\partial n_F(\epsilon)}{\partial \epsilon}
\left\{\left[\delta\mu(\hat{\mathbf{v}})-\frac{\mu}{T}\delta T(\hat{\mathbf{v}})\right]
+\frac{\epsilon}{T}\delta T(\hat{\mathbf{v}})\right\}.
\end{eqnarray}

The Ansatz Eq.~(\ref{EqAnsatz}) greatly simplifies the solution of the kinetic equation, replacing the integral equation by a matrix one. 
Furthermore, the same kinematics-induced singularity in the collinear scattering as in $I_{ii}$ appears in
also the intralayer collision integrals $I_{ij}$. This implies fast unidirectional thermalization between the layers.
Therefore, we set $\chi_T^{(1)}=\chi_T^{(2)}$, and hence reduce the kinetic equation for the double-layer setup to a
$3\times3$ matrix equation (``three-mode approximation'').

\subsection{D. Hydrodynamic equations}

In each of the two layers, we introduce the particle currents (the total velocities)
\begin{equation}
\mathbf{J}_i = - N T \int d\epsilon\ \nu(\epsilon)  \frac{\partial n_F^{(i)}}{\partial \epsilon} \int d\hat{\mathbf{v}}\  \mathbf{v} \,
h_\alpha(\epsilon,\mathbf{v}),
\end{equation}
and energy currents (or, equivalently, the total momenta)
\begin{equation}
\mathbf{P}_i = - N \int d\epsilon\ \nu(\epsilon)\ \epsilon \ \frac{\partial n_F^{(i)}}{\partial \epsilon} \int d\hat{\mathbf{v}}\ \mathbf{v} \,
h_i(\epsilon,\mathbf{v}),
\end{equation}
and substitute the Ansatz, Eq. \ref{EqAnsatz}, into these expressions,
which yields
\begin{eqnarray}
\mathbf{J}_i &=& \frac{N}{2} \left[ A_0^{(i)} \chi_0^{(i)} + A_1^{(i)} \chi_1^{(i)}\right] \mathbf{E} \\
\mathbf{P}_i &=& \frac{N}{2} \left[ A_1^{(i)} \chi_0^{(i)} + A_2^{(i)} \chi_1^{(i)}\right] \mathbf{E}.
\end{eqnarray}
Here
\begin{equation}
A_n^\alpha= -\frac{v^2}{T} \int d\epsilon\ \nu(\epsilon)\ \frac{\partial n_F^\alpha}{\partial \epsilon}\ \left(\frac{\epsilon}{T}\right)^n.
\end{equation}

In terms of the currents, the fast interlayer thermalization ($\chi_T^{(1)}=\chi_T^{(2)}$) 
translates into the relation
\begin{equation}
B_2^b(\mathbf{P}_a-B_1^a \mathbf{J}_a)=B_2^a(\mathbf{P}_b-B_1^b \mathbf{J}_b),
\end{equation}
where
\begin{eqnarray} \label{EqDefiningTheBs}
B_0^\alpha = A_0^\alpha,\qquad B_1^\alpha=\frac{A_1^\alpha}{A_0^\alpha},\qquad
B_2^\alpha = A_2^\alpha-\frac{(A_1^\alpha)^2}{A_0^\alpha}.
\end{eqnarray}
The asymptotics of the functions $B^{(i)}_n(\mu_i/2T)$ read:
\begin{equation} \label{EqAssymptoticsOfBs}
B^{(i)}_n(x\ll 1)=\left\{\begin{array}{cc}
         \ln2/\pi & n=0\\
	 4x & n=1\\
	 9\zeta(3)/2\pi & n=2
        \end{array} \right. , \qquad  B_n(x\gg 1)=\left\{\begin{array}{cc}
         |x|/\pi & n=0\\
	 2x& n=1\\
	 \pi |x|/3 & n=2
        \end{array} \right. .
\end{equation}
It is also convenient to define
$
B_2=B_2^{(1)}+B_2^{(2)}.
$
Finally, we introduce the total momentum (total energy current)
\begin{equation}
\mathbf{P}=\mathbf{P}_a+\mathbf{P}_b,
\end{equation}
which is not affected by electron-electron collisions due to total momentum conservation
in the e-e collision integral.

These transformations allow us to rewrite the matrix kinetic equation in the ``hydrodynamic form''.
Here we can also introduce electric fields in both layers without doubling the number of relevant modes.
Integrating the reduced matrix kinetic equation over the energy  
with
\begin{eqnarray}
&&-NT\int d\epsilon_1 \nu(\epsilon)\frac{\partial n_F(\epsilon_1)}{\partial \epsilon_1}
\{\ldots\}
=N\int d\epsilon_1 \frac{\nu(\epsilon)}{\cosh^2\frac{\epsilon_1-\mu_i}{2T}}
\{\ldots\}
\\
&&\text{and}\nonumber
\\
&&-N\int d\epsilon_1 \epsilon_1 \nu(\epsilon)\frac{\partial n_F(\epsilon_1)}{\partial \epsilon_1}
\{\ldots\}
=N\int d\epsilon_1 \epsilon_1\frac{\nu(\epsilon)}{\cosh^2\frac{\epsilon_1-\mu_i}{2T}}
\{\ldots\}
\end{eqnarray}
yields the following steady-state equations (in order to avoid confusion in indices, from now on we label the layers by $a$ and $b$)
\begin{eqnarray}
\left\{\frac{1}{\tau} + \left[(B_1^a)^2+\frac{B_2}{B_0^a}\right] \frac{1}{\tau^a_{ee}} - B_1^a B_1^b  \frac{1}{\tau_{D}} \right\}\mathbf{J}_a
&+&\left\{ B_1^a B_1^b  \frac{1}{\tau^a_{ee}}-\left[(B_1^b)^2+\frac{B_2}{B_0^b}\right] \frac{1}{\tau_{D}} \right\}\mathbf{J}_b\nonumber \\
&=&\frac{2}{N} T B_0^a e\mathbf{E}_a+\left( \frac{B_1^a}{\tau^a_{ee}} -  \frac{B_1^b}{\tau_{D}}\right)\mathbf{P}
\\
\left\{\frac{1}{\tau} + \left[(B_1^b)^2+\frac{B_2}{B_0^b}\right] \frac{1}{\tau^b_{ee}} - B_1^a B_1^b  \frac{1}{\tau_{D}} \right\}\mathbf{J}_b
&+&\left\{ B_1^a B_1^b  \frac{1}{\tau^b_{ee}}- \left[(B_1^a)^2+\frac{B_2}{B_0^a}\right] \frac{1}{\tau_{D}} \right\}\mathbf{J}_a\nonumber \\
&=&\frac{2}{N} T B_0^b e\mathbf{E}_b+\left( \frac{B_1^b}{\tau^b_{ee}}-  \frac{B_1^a}{\tau_{D}}\right)\mathbf{P}
\\
\frac{\mathbf{P}}{\tau}&=& \frac{2}{N} T \left(B_0^a B_1^a \mathbf{E}_a+B_0^b B_1^b \mathbf{E}_b\right).
\label{hydro-drag}
\end{eqnarray}
Here we have introduced the following effective transport relaxation rates:
\begin{eqnarray}
\frac{1}{\tau^a_{ee}}&=&\frac{N}{8T^2B_2} 
\int\mathrm{d}\{\epsilon_i\}\mathrm{d}\{\hat{\mathbf{v}}_i\}\
\left[ 
\left(\mathbf{v}_1-\mathbf{v}_2+\mathbf{v}_3-\mathbf{v}_4\right)^2\ \mathcal{W}^{aa}\ +
2  \left(\mathbf{v}_1-\mathbf{v}_2\right)^2\ \mathcal{W}^{ab} \right],
\\
\frac{1}{\tau^b_{ee}}&=&\frac{N}{8T^2B_2} 
\int\mathrm{d}\{\epsilon_i\}\mathrm{d}\{\hat{\mathbf{v}}_i\}\ 
\left[ \left(\mathbf{v}_1-\mathbf{v}_2+\mathbf{v}_3-\mathbf{v}_4\right)^2\ \mathcal{W}^{bb}\ +
2 \left(\mathbf{v}_1-\mathbf{v}_2\right)^2 \mathcal{W}^{ba} \right],
\\
\frac{1}{\tau_D}&=&\frac{N}{4T^2B_2} 
\int\mathrm{d}\{\epsilon_i\}\mathrm{d}\{\hat{\mathbf{v}}_i\}\  
 \left(\mathbf{v}_1-\mathbf{v}_2\right)\left(\mathbf{v}_4-\mathbf{v}_3\right)\ \mathcal{W}^{ba}.
 \label{rates}
\end{eqnarray}
Here $1/\tau^{a(b)}_{ee}$ are the intralayer transport relaxation rates describing the velocity relaxation
within a layer due to inelastic scattering with electrons in the same layer (described by $\mathcal{W}^{aa}$) and 
in the other layer ($\mathcal{W}^{ab}$ term). The rate $1/\tau_D$ describes the interlayer velocity relaxation (velocity transfer from one layer to the other due to $\mathcal{W}^{ab}$): we call it the drag rate. 
The kernels $\mathcal{W}^{ij}$ here are related to the kernel of the collision integral (\ref{eq4}) as follows:
\begin{equation}
\mathcal{W}^{ij}(1,2;3,4)=\frac{\nu(\epsilon_1)}{\cosh^2\frac{\epsilon_1-\mu_i}{2T}}W^{ij}(1,2;3,4).
\end{equation}

\section{2. Ballistic regime}

\subsection{A. Resistivity matrix}

The hydrodynamic equations (\ref{hydro-drag}) yield the following explicit expressions for the intralayer and 
interlayer resistivities:
\begin{eqnarray}
\rho_{aa}&=&\frac{\hbar}{e^2}\frac{2 B_2}{N \left(B_0^a\right)^2 T}
\left[\frac{B_0^a}{\tau}+
\dfrac{\dfrac{1}{\tau\tau^a_{ee}}+\left(B_1^b\right)^2\left(\dfrac{1}{\tau^a_{ee}\tau^b_{ee}}-\dfrac{1}{\tau_D^2}\right)}
{\dfrac{1}{\tau}+\dfrac{\left(B_1^a\right)^2}{\tau^a_{ee}}
+\dfrac{\left(B_1^b\right)^2}{\tau^b_{ee}}
-\dfrac{2B_1^aB_1^b}{\tau_D}}\right],
\label{rhoaa}
\\
\rho_{ab}&=&-\frac{\hbar}{e^2}\frac{2 B_2}{N  B_0^a B_0^b T}
\dfrac{\dfrac{1}{\tau\tau_D}+B_1^aB_1^b\left(\dfrac{1}{\tau^a_{ee}\tau^b_{ee}}-\dfrac{1}{\tau_D^2}\right)}
{\dfrac{1}{\tau}+\dfrac{\left(B_1^a\right)^2}{\tau^a_{ee}}
+\dfrac{\left(B_1^b\right)^2}{\tau^b_{ee}}
-\dfrac{2B_1^aB_1^b}{\tau_D}} 
\label{rhoab}
\end{eqnarray}
$\rho_{bb}=\rho_{aa}(a\leftrightarrow b)$, and $\rho_{ba}=\rho_{ab}$.
The drag coefficient is defined as
\begin{equation}
\rho_D=-\rho_{ab}.
\end{equation}
It can also be rewritten in the following form,
\begin{equation}
\rho_D=\frac{\hbar}{e^2}\frac{2 B_2}{N  B_0^a B_0^b T \tau_D}
\left[1-\dfrac{\left(\dfrac{B_1^a}{\tau^a_{ee}}+\dfrac{B_1^b}{\tau_D}\right)
\left(\dfrac{B_1^b}{\tau^b_{ee}}+\dfrac{B_1^a}{\tau_D}\right)}
{\dfrac{1}{\tau_D}\left(\dfrac{1}{\tau}+\dfrac{\left(B_1^a\right)^2}{\tau^a_{ee}}
+\dfrac{\left(B_1^b\right)^2}{\tau^b_{ee}}
-\dfrac{2B_1^aB_1^b}{\tau_D}\right)} \right],
\end{equation}
which for the disorder-dominated case yields directly the conventional
perturbative drag:
\begin{equation}
\rho_D=\frac{\hbar}{e^2}\frac{2 B_2}{N  B_0^a B_0^b T \tau_D}.
\end{equation}

For equal layers, we denote
\begin{equation}
\epsilon_0=2B_0 T/N, \qquad C_1=B_1, \qquad C_2=B_2/B_0.
\end{equation}
Then 
the resistivity matrix reads
\begin{eqnarray}
{\hat \rho} = \frac{\hbar}{e^2}\frac{C_2}{\epsilon_0}
\left\{
\frac{1}{\tau} \left[
\begin{pmatrix}
 1 &  0 \\ 
 0 & 1
  \end{pmatrix}
  \right.\right.
  &+&
  \left.
  \dfrac{C_2}{1/\tau+2 C_1^2\left(1/\tau_{ee}-1/\tau_D\right)}
\begin{pmatrix}
 1/\tau_{ee} &  -1/\tau_D \\[0.5pt]
 -1/\tau_D & 1/\tau_{ee}
  \end{pmatrix}\right]
  \nonumber 
  \\[1pt]
  &+&
  \left.
\dfrac{C_2 C_1^2\left(1/\tau_{ee}^2-1/\tau_D^2\right)}
{1/\tau+2 C_1^2\left(1/\tau_{ee}-1/\tau_D\right)}
\begin{pmatrix}
 1 &  -1 \\ -1 & 1
  \end{pmatrix}
  \right\}.
  \label{res-matrix}
\end{eqnarray}
Here the first term is the intralayer resistivity determined by disorder, the second term describes the conventional Coulomb drag in combination with the intralayer inelastic transport relaxation, and the last term
arises due to the fast unidirectional thermalization in graphene.

In the clean limit $\tau=\infty$ the resistivity matrix has the form
\begin{equation}
{\hat \rho} = \frac{\hbar}{e^2}\frac{2 B_2}{N T}
\dfrac{\dfrac{1}{\tau^a_{ee}\tau^b_{ee}}-\dfrac{1}{\tau_D^2}}
{\dfrac{\left(B_1^a\right)^2}{\tau^a_{ee}}
+\dfrac{\left(B_1^b\right)^2}{\tau^b_{ee}}
-\dfrac{2B_1^aB_1^b}{\tau_D}}
\begin{pmatrix}
 \dfrac{(B_1^b)^2}{(B_0^a)^2} &  -\dfrac{B_1^a B_1^b }{B_0^a B_0^b} \\ -\dfrac{B_1^a B_1^b }{B_0^a B_0^b} & \dfrac{(B_1^a)^2}{(B_0^b)^2},
  \end{pmatrix}
  \label{rho-clean}
\end{equation}
which for equal layers simplifies to 
\begin{equation}
{\hat \rho} = \frac{\hbar}{e^2}
\frac{C_2}{\epsilon_0}\ \left(\frac{1}{\tau_{ee}}+\frac{1}{\tau_D}\right)
\displaystyle\begin{pmatrix}
 1 &  -  1 \\ - 1 & 1
  \end{pmatrix}.
  \label{rho-clean-equal}
\end{equation}
The off-diagonal component of this matrix is given in Eq.~(14) of the main text.

\subsection{B. Asymptotics of the drag coefficient}

The general condition separating the disorder-dominated and Coulomb-dominated transport regimes
can be found from Eq.~(\ref{rhoab}), where one should compare the two terms in the numerator:
\begin{equation}
\dfrac{1}{\tau} \sim B_1^aB_1^b \tau_D\left(\dfrac{1}{\tau^a_{ee}\tau^b_{ee}}-\dfrac{1}{\tau_D^2}\right).
\label{condition}
\end{equation}
In the vicinity of the Dirac point ($\mu_{a,b}\ll T$), the intralayer transport relaxation rates are
\begin{equation}
\frac{1}{\tau^{a,b}_{ee}}\sim \alpha^2 N T,
\end{equation}
whereas the drag rate was found in Ref. \cite{us1}:
\begin{equation}
\frac{1}{\tau_D}\sim \alpha^2 N \frac{\mu_a\mu_b}{T},
\end{equation}
so that $1/\tau_{ee}\gg 1/\tau_D$.
Substituting these results into Eq. (\ref{condition}),
we find
\begin{equation}
\dfrac{1}{\tau} \sim \frac{\mu_a\mu_b}{T^2} \frac{\tau_D}{\tau^a_{ee}\tau^b_{ee}}
\sim \alpha^2 N T,
\end{equation}
i.e. the crossover occurs at $1/\tau\sim 1/\tau_{ee}$.

Away from the Dirac point, $T\ll \mu_{a,b}\ll T/\alpha$, the drag rate is (for simplicity we set $\mu_a\sim \mu_b\sim \mu$)
\begin{equation}
\frac{1}{\tau_D}\sim \alpha^2 N \frac{T^2}{\mu} \ln\frac{\mu}{T},
\end{equation}
and
\begin{equation}
\frac{1}{\tau_{ee}}-\frac{1}{\tau_D}\sim \frac{1}{\tau_{ee}}\frac{T^2}{\mu^2}\ll \frac{1}{\tau_{ee}}.
\end{equation}
It is worth mentioning that, for $\mu\gg T$, the intralayer scattering is much less efficient for the
relaxation of velocity than the interlayer scattering. Indeed, the intralayer 
($\propto \mathcal{W}^{ii}$) contribution to $1/\tau_{ee}$ in Eq. (\ref{rates}) contains the combination of velocities $\mathbf{v}_1-\mathbf{v}_2+\mathbf{v}_3-\mathbf{v}_4$, which for $\mu\gg T$ is very close to
 $\mathbf{p}_1-\mathbf{p}_2+\mathbf{p}_3-\mathbf{p}_4$. Therefore, at high chemical potentials
 $1/\tau_{ee}$ is dominated by the interlayer contribution ($\propto \mathcal{W}^{ab}$).
Thus the crossover between the two regimes occurs for $\mu\gg T$ at
\begin{equation}
\dfrac{1}{\tau} \sim \frac{\mu^2}{T^2}\left(\dfrac{1}{\tau_{ee}}-\dfrac{1}{\tau_D}\right)
\sim \alpha^2 N \frac{T^2}{\mu} \ln\frac{\mu}{T} \sim \frac{1}{\tau_{ee}}.
\label{condition-high-mu}
\end{equation}
The drag coefficient in the disorder-dominated regime coincides
with the perturbative result,
\begin{eqnarray}
\rho_D &=& \frac{\hbar}{e^2}
\frac{2 B_2}{N B_0^a B_0^b T \tau_D}\nonumber \\
 &\sim& \frac{\hbar}{e^2}\ \alpha^2 \frac{T^2(\mu_a+\mu_b)}{(\mu_a\mu_b)^2}
 \ln\frac{\text{min}\{\mu_a,\mu_b\}}{T}, \qquad T\ll \mu \ll \frac{T}{\alpha},\ \frac{1}{\tau}\gg \alpha^2 N 
 \frac{T^2}{\mu}.
\end{eqnarray}
In the opposite, ultraclean case, assuming for simplicity equal layers, we
get from Eq.~(\ref{rho-clean-equal})  
\begin{eqnarray}
\rho_D&=&\frac{\hbar}{e^2}
\frac{C_2}{\epsilon_0}\ \left(\frac{1}{\tau_{ee}}+\frac{1}{\tau_D}\right)
\simeq \frac{\hbar}{e^2} \frac{2C_2}{\epsilon_0 \tau_D}\nonumber \\
&\sim& \frac{\hbar}{e^2}\ \alpha^2 \frac{T^2}{\mu^2}
 \ln\frac{\mu}{T}, \qquad\qquad T\ll \mu \ll \frac{T}{\alpha},\ \frac{1}{\tau}\ll \alpha^2 N
 \frac{T^2}{\mu}.
\end{eqnarray}
The difference between the disordered (perturbative) and ultraclean (equilibrated)
results for the drag is in the presence of $1/\tau_{ee}$ in the latter case.
In particular, for equal layers, this enhances the drag by a factor of 2.
Since these results are qualitatively the same, we do not analyze the behavior of the drag
at yet higher chemical potentials, referring the reader to Ref. \cite{us1}.

\subsection{C. Third-order drag rate}

It is important that the second-order (Golden-rule) drag rate vanishes at the Dirac point
due to the particle-hole symmetry. However, the particle-hole symmetry does not affect
the odd-order drag rates \cite{lak}. Near the Dirac point, such rates should not depend on $\mu$ and hence
are proportional to $T$. Taking into account the second-order matrix element $M_{ab}^{(2)}\propto \alpha^2$,
\begin{equation}
\left|M_{ab}^{(1)}+M_{ab}^{(2)}\right|^2\simeq \left|M_{ab}^{(1)}\right|^2
+ 2 \text{Re} \left\{M_{ab}^{(1)} \left[M_{ab}^{(2)}\right]^*\right\},
\end{equation}
we estimate
\begin{equation}
\frac{1}{\tau_D} \sim \alpha^2 N \frac{\mu^2}{T^2} + \alpha^3 N T,
\end{equation}
where we skip the numerical prefactors in both terms.
A similar correction arises in $1/\tau_{ee}$, but there it is always subleading for $\alpha\ll 1$.
Substituting this correction into Eq. (\ref{res-matrix}), we find
\begin{equation}
\rho_D\sim \frac{\hbar}{e^2}\frac{1}{NT}
\dfrac{\dfrac{N}{\tau}\left(\alpha^2 \dfrac{\mu^2}{T^2} + \alpha^3 T\right)+\dfrac{\mu^2 N^2}{T^2}\left[\alpha^4 T^2-\left(\alpha^2 \dfrac{\mu^2}{T^2} + \alpha^3 T\right)^2\right]}
{\dfrac{1}{\tau}+\dfrac{\mu^2 N}{T^2} \left[\alpha^2 T + \left(\alpha^2 \dfrac{\mu^2}{T^2} + \alpha^3 T\right) \right]} 
\end{equation}
Clearly, for $\mu\gg \alpha^{1/2} T$ we can disregard the $\alpha^3$-terms.
In the opposite limit, we neglect the conventional drag term:
\begin{equation}
\rho_D \sim \frac{\hbar}{e^2}
\dfrac{\alpha^3 T+\alpha^4 \mu^2 \tau N}
{T+\alpha^2 \mu^2\tau N}, \qquad \mu\ll\alpha^{1/2} T, \quad \frac{1}{\tau}\ll T.
\end{equation}
Exactly at  the Dirac point this yields 
\begin{equation}
\rho_D\sim \frac{\hbar}{e^2} \alpha^3.
\label{alpha3}
\end{equation}
For finite chemical potential, this result is valid for
\begin{equation}
\frac{1}{\tau}\gg \alpha N \frac{\mu^2}{T}.
\end{equation}
In the opposite limit, one can neglect the $\alpha^3$-term,
yielding
\begin{equation}
\rho_D \sim \frac{\hbar}{e^2}
\dfrac{\alpha^4 \mu^2 \tau N}
{T+\alpha^2 \mu^2\tau N} = \frac{\hbar}{e^2}\begin{cases}
\alpha^2, &\quad 1/\tau\ll \alpha^2 N \mu^2/T \\
\alpha^4 N \mu^2 \tau/T, &\quad \alpha^2 N \mu^2/T\ll 1/\tau\ll \alpha N \mu^2/T
\end{cases}.
\end{equation}

\section{3. Diffusive regime}

\subsection{A. Third-order contribution to the drag}

In this section we analyze the third-order drag \cite{lak} in the
diffusive regime $T\tau\ll 1$ for the case of high dimensionless
conductances (per spin and per valley), $g=\nu D \sim \mu\tau \gg 1$,
where $D=v^2\tau/2$ is the diffusion coefficient. In this limit, one
can calculate the prefactor analytically.  In the vicinity of the
Dirac point $\mu\tau\ll 1$, the conductance is of order unity.
Indeed, experiments on high-quality samples show $T$-independent
$\sigma$ down to $T=30$ mK \cite{kim}, that can be explained by the
specific character of disorder in graphene \cite{ogm}.  Therefore,
there we also assume the diffusive dynamics described by the diffusion
propagators
\begin{equation}
\mathcal{D}_i(q,\omega)=\frac{1}{\nu_{i}}\, \frac{1}{D_{i}q^{2}-i\omega}, \qquad qv,\omega \ll 1/\tau,
\end{equation}
and polarization operators
\begin{equation}
\Pi_i(q,\omega)=N\nu_i\dfrac{D_{i}q^{2}}
{D_{i}q^{2}-i\omega},
\end{equation}
where $\nu_{i}$ is the density of states of the layer (per spin and valley)
$i=1,2$  and $D_i$ is its diffusion coefficient. 

The third-order drag resistivity was calculated in Ref. \cite{lak} for the case of
large interlayer distance, $\kappa d\gg 1$. Here we generalize this result to the opposite case
$\kappa d\ll 1$, which is experimentally relevant for graphene near the Dirac point.

The analytic expression for the third-order drag resistivity is given by \cite{lak}
\begin{eqnarray}
\rho_{D}^{(3)}&=&\frac{\hbar}{e^2}\ 32 T g_{1}g_{2}\int\limits^{\infty}_{0}
\mathrm{d}\omega\mathrm{d}\Omega\,
\mathcal{F}_{1}(\omega,\Omega)\mathcal{F}_{2}(\omega,\Omega)\nonumber \\
&\times& \int \frac{d^2q}{(2\pi)^2} \int \frac{d^2Q}{(2\pi)^2}
\, \mathrm{Im}\Big[\mathcal{D}_{1}(q,\omega)\mathcal{D}_{2}(q,\omega)
 \mathcal{V}_{12}(q,\omega)
\mathcal{V}_{12}\left(\frac{\mathbf{q}}{2}-\mathbf{Q},{\frac{\omega}{2}}-\Omega\right)
\mathcal{V}_{12}\left(\frac{\mathbf{q}}{2}+\mathbf{Q},{\frac{\omega}{2}}+\Omega\right)\Big].
\nonumber
\\
\label{sigma-general}
\end{eqnarray}
Here the thermal factors $\mathcal{F}_{i}(\omega,\Omega)$ are
given by
\begin{subequations}\label{Spectral-F}
\begin{equation}
\mathcal{F}_{1}(\omega,\Omega)=T\frac{\partial}{\partial\Omega}
\left[\mathcal{B}(\Omega+\omega/2)-\mathcal{B}(\Omega-\omega/2)\right],
\end{equation}
\begin{equation}
\mathcal{F}_{2}(\omega,\Omega)=2-\mathcal{B}(\Omega+\omega/2)
-\mathcal{B}(\Omega-\omega/2)+\mathcal{B}(\omega),
\end{equation}
where
\begin{equation}
\mathcal{B}(\omega)=\frac{\omega}{T}\coth\left(\frac{\omega}{2T}\right).
\end{equation}
\end{subequations}
The propagators of longitudinal vector potentials $\mathcal{V}_{12}(q,\omega)$ in Eq. (\ref{sigma-general})
include the dressing of the vertices by diffusons:
\begin{equation}
\mathcal{V}_{12}(q,\omega)=
\frac{q^{2}U^{\text{RPA}}_{12}(q,\omega)}
{(D_{1}q^{2}-i\omega)(D_{2}q^{2}-i\omega)},
\label{calV}
\end{equation}
where the retarded RPA-screened interlayer interaction $U^{\text{RPA}}_{12}(q,\omega)$ is
defined in Eq.~(\ref{RPA}).

For simplicity, we consider equal layers.
For small interlayer distance $d\ll v\tau$ we have $qd\ll 1$. 
It is convenient to introduce the inverse screening length
\begin{equation}
\kappa=2\pi e^2 \nu,
\end{equation}
where $\nu$ is the thermodynamic density of states per one flavor of particles.
Expanding $\exp(-qd)\simeq 1-qd$ we get
\begin{eqnarray}
U^{\text{RPA}}_{12}(q,\omega)
&=&
 U_{12}^{(0)}(q)
\left[\left(1+U_{11}^{(0)}(q) \Pi(q,\omega)\right)^2-\left( U_{12}^{(0)}(q) \Pi(q,\omega)\right)^2\right]^{-1}
\nonumber \\
&=&\frac{1}{\nu} \frac{\kappa e^{-qd}}{q}
\left[\left(1+\frac{N\kappa}{q}\frac{Dq^2}{Dq^2-i\omega}\right)^2-\left(\frac{N\kappa e^{-qd}}{q}\frac{Dq^2}{Dq^2-i\omega}\right)^2\right]^{-1}
\nonumber \\
&\simeq&\frac{1}{\nu} \frac{\kappa}{q}\frac{(Dq^2-i\omega)^2}{[Dq(q+2N\kappa)-i\omega] [Dq^2(1+N\kappa d) -i\omega]},
\end{eqnarray}
and hence
\begin{equation}
\mathcal{V}_{12}(q,\omega)=
\frac{1}{\nu} \frac{\kappa q}{[Dq(q+2N\kappa)-i\omega] [Dq^2(1+N\kappa d) -i\omega]}.
\label{ourV12}
\end{equation}

We first consider the case of large interlayer separation, $N\kappa d\gg 1$.
For $N\kappa \gg \text{max}\{1/d,(T/D)^{1/2}\}$ we find
\begin{equation}
\mathcal{V}_{12}(q,\omega)\simeq
\frac{1}{2 N g} \frac{1}{Dq^2 N\kappa d -i\omega},
\end{equation}
which reproduces the result of Ref. \cite{lak}:
\begin{equation}
\rho_D^{(3)}\sim \frac{\hbar}{e^2}\, \frac{1}{N^3 g^3}\, \frac{1}{(N\kappa d)^2}.
\end{equation}
For $1/d\ll N\kappa \ll (T/D)^{1/2}$,
\begin{equation}
\mathcal{V}_{12}(q,\omega)\simeq
\frac{1}{\nu}\frac{\kappa q}{Dq^2-i\omega} \frac{1}{Dq^2 N\kappa d -i\omega},
\end{equation}
and we find
\begin{equation}
\rho_D^{(3)}\sim \frac{\hbar}{e^2}\, \frac{1}{g^3}\, \frac{1}{(N\kappa d)^2}  \left(\frac{D \kappa^2}{T}\right)^{3/2}.
\end{equation}

In the opposite case $N\kappa d\ll 1$ (which is relevant to our problem),
\begin{equation}
\mathcal{V}_{12}(q,\omega)\simeq
\frac{1}{\nu} \frac{\kappa q}{[Dq(q+2N\kappa)-i\omega]\ [Dq^2-i\omega]}.
\end{equation}
Substituting this into Eq. (\ref{sigma-general}), we get
\begin{eqnarray}
 \rho_{D}^{(3)}\!&=&\!\frac{\hbar}{e^2}\, 32T\frac{g^2}{\nu^5}\int\limits^{\infty}_{0}
\mathrm{d}\omega\mathrm{d}\Omega\,
\mathcal{F}_{1}(\omega,\Omega)\mathcal{F}_{2}(\omega,\Omega)\!
\int \frac{d^2q}{(2\pi)^2} \int \frac{d^2Q}{(2\pi)^2}\ \mathrm{Im}\left\{\Big[\frac{1}{Dq^2-i\omega}\Big]^2\
\frac{\kappa q}{[Dq(q+2 N \kappa)-i\omega][Dq^2-i\omega]} \right.
\nonumber \\
&&
\nonumber
\\
&\times&
\left.
\frac{\kappa |\mathbf{q}/2-\mathbf{Q}|}{[D(\mathbf{q}/2-\mathbf{Q})^2+2 D|\mathbf{q}/2-\mathbf{Q}| N \kappa-i(\omega/2-\Omega)]\ [D(\mathbf{q}/2-\mathbf{Q})^2-i(\omega/2-\Omega)]}
\right.
\nonumber \\
&&
\nonumber
\\
&\times&
\left.
\frac{\kappa |\mathbf{q}/2-\mathbf{Q}|}{[D(\mathbf{q}/2+\mathbf{Q})^2+2 D|\mathbf{q}/2+\mathbf{Q}| N \kappa-i(\omega/2+\Omega)]\ [D(\mathbf{q}/2+\mathbf{Q})^2-i(\omega/2+\Omega)]}
\right\}.
\label{sigma-small}
\end{eqnarray}
The frequency integrals are dominated by $\omega\sim\Omega\sim T$, whereas the momentum integrals
are dominated by $q\sim Q\sim q_T=\sqrt{T/D}$. Therefore, the drag conductivity in Eq.~(\ref{sigma-small})
can be estimated as
\begin{eqnarray}
\rho_D^{(3)}
&\sim &\frac{\hbar}{e^2} \frac{T g^2}{\nu^5}\ \underbrace{T^2}_{d\omega d\Omega} \ \underbrace{q_T^4}_{d^2q d^2 Q} \ \frac{1}{(Dq_T^2+T)^5}\frac{\kappa^3 q_T^3}{(Dq_T^2+T+Dq_TN\kappa)^3}
\nonumber \\
&&
\nonumber \\
&\sim &\frac{\hbar}{e^2} \frac{1}{g^3}\ 
\left(\frac{\kappa}{N\kappa+\sqrt{T/D}}\right)^3.
\end{eqnarray}
Therefore, for $N \kappa\gg\sqrt{T/D}$ we get
\begin{equation}
\rho_D^{(3)} \sim \frac{\hbar}{e^2}\ \frac{1}{N^3 g^3},
\end{equation}
while for $N \kappa\ll\sqrt{T/D}$ we find
\begin{equation}
\rho_D^{(3)} \sim \frac{\hbar}{e^2}\ \frac{1}{g^3} \ \left(\frac{D \kappa^2}{T}\right)^{3/2}.
\end{equation}

In the first case the expression for the third-order drag coefficient is given by
\begin{eqnarray}
 \rho_{D}^{(3)}\!&=&\!\frac{\hbar}{e^2}\ 32T\frac{g^2}{\nu^5}\int\limits^{\infty}_{0}
\mathrm{d}\omega\mathrm{d}\Omega\,
\mathcal{F}_{1}(\omega,\Omega)\mathcal{F}_{2}(\omega,\Omega)\!
\int \frac{d^2q}{(2\pi)^2} \int \frac{d^2Q}{(2\pi)^2}\
 \mathrm{Im}\left\{\Big[\frac{1}{Dq^2-i\omega}\Big]^2\
\right.
\nonumber \\
&&
\nonumber
\\
&\times&
\left.
\left(\frac{1}{2 N D}\right)^3
\frac{1}{Dq^2-i\omega}
\frac{1}{D(\mathbf{q}/2-\mathbf{Q})^2-i(\omega/2-\Omega)}
\
\frac{1}{D(\mathbf{q}/2+\mathbf{Q})^2-i(\omega/2+\Omega)}
\right\}.
\label{sigma-small-1}
\end{eqnarray}
The integrals here are now dimensionless [one measures momenta in
  units of $(T/D)^{1/2}$ and frequencies in units of $T$].  In the second case
the prefactor is again determined by a dimensionless integral:
\begin{eqnarray}
 \rho_{D}^{(3)}\!&=&\!\frac{\hbar}{e^2}\ 32T\frac{g^2}{\nu^5}  \ \int\limits^{\infty}_{0}
\mathrm{d}\omega\mathrm{d}\Omega \,
\mathcal{F}_{1}(\omega,\Omega)\mathcal{F}_{2}(\omega,\Omega)\!
\int \frac{d^2q}{(2\pi)^2} \int \frac{d^2Q}{(2\pi)^2}\  \mathrm{Im}\left\{\Big[\frac{1}{Dq^2-i\omega}\Big]^2\
\right.
\nonumber \\
&&
\nonumber
\\
&\times&
\left. \kappa^3
\frac{q}{[Dq^2-i\omega]^2}
\frac{|\mathbf{q}/2-\mathbf{Q}|}{[D(\mathbf{q}/2-\mathbf{Q})^2-i(\omega/2-\Omega)]^2}
\
\frac{|\mathbf{q}/2+\mathbf{Q}|}{[D(\mathbf{q}/2+\mathbf{Q})^2-i(\omega/2+\Omega)]^2}
\right\}.
\label{sigma-small-2}
\end{eqnarray}

For $T\tau\ll 1$ and $\mu\tau\gg 1$, we have 
\begin{equation}
\kappa\sim \alpha \mu/v, 
\end{equation}
which implies
\begin{equation}
N\kappa=\sqrt{\frac{T}{D}}\quad \leftrightarrow \quad \frac{1}{\tau}=\frac{\alpha^2N^2\mu^2}{T}.
\end{equation}
For $\mu\tau \ll 1$, we have 
\begin{equation}
\kappa\sim \alpha/v\tau,
\end{equation}
and hence 
\begin{equation}
N\kappa=\sqrt{\frac{T}{D}}\quad \leftrightarrow \quad \frac{1}{\tau}=\frac{T}{\alpha^2N^2}.
\end{equation}
Thus, for 
$$\text{max}\{T,\alpha^2N^2\mu^2/T\}\ll 1/\tau \ll T/\alpha^2 N^2$$ the third-order drag
resistivity reads:
\begin{equation}
\rho_D^{(3)}\sim \frac{\alpha^3}{(T\tau)^{3/2}}.
\end{equation}
At $T\tau\sim 1$ this result matches the ballistic third-order drag resistivity, Eq. (\ref{alpha3}).

For yet lower $T\ll \alpha^2N^2/\tau$ and $\mu\tau\ll 1$ the third-order drag
saturates at
\begin{equation}
\rho_D^{(3)}\sim \frac{\hbar}{e^2}\,\frac{1}{N^3}.
\end{equation}
Finally, for $\mu\tau\gg \text{max}\{1,(T\tau)^{1/2}/\alpha N\}$,
the third-order drag behaves as
\begin{equation}
\rho_D^{(3)}\sim \frac{\hbar}{e^2}\,\frac{1}{(N \mu \tau)^3}.
\end{equation}

\section{4. Correlated disorder}

In the original version of the paper we mentioned the correlations
between the disorder potentials of the two layers \cite{gor} as an
alternative mechanism leading to a low-$T$ peak in $\rho_D$ at the
Dirac point.  After the submission of the original version, we became
aware of a preprint by Song and Levitov \cite{lev} that focused on
such a mechanism.  In this section we analyze the role of interlayer
correlations of disorder potentials (both of short-range and
long-range nature) within our general framework. This allows us to
compare the effect of correlated disorder with the third-order drag
considered in the main text and in Sections 2C and 3 of the
Supplemental Material.
 
As emphasized in Ref.~\cite{lev}, the correlations between the
disorder potentials of the two layers might be especially important in
drag experiments on graphene near the Dirac point for the two reasons:
(i) similarly to the third-order drag, it does not require \cite{gor}
the particle-hole symmetry and hence provides finite drag at the
charge neutrality point \cite{lev}; (ii) in contrast to experiments on
conventional semiconducting double wells, the interlayer distance in
graphene experiments is rather small, which enhances the disorder
correlations between the layers. In what follows, we analyze the two
models of correlations: (A) Correlated scattering off common
short-range impurities \cite{gor} and (B) correlations of large-scale
inhomogeneities of the chemical potentials in the layers \cite{lev}.

\subsection{A. Short-range correlations: correlated impurity scattering}

Following Ref. \cite{gor}, we introduce the matrix of disorder correlators
$w^{(ij)}_{{\hat{\mathbf v}} {\hat{\mathbf v}}'}=\langle u^{(i)} u^{(j)} \rangle_{\text{imp}}$.
The values of $w^{(ij)}$ at $i \neq j$ differ
from zero due to correlations between the impurity
potentials $u^{(i)}$ in different layers. The 
total scattering rates are
defined by 
\begin{eqnarray}
\frac{1}{\tau_{ij}}&=&\left<w^{ij}_{{\hat{\mathbf v}} {\hat{\mathbf v}}'}\frac{1+\hat{\mathbf{v}}\hat{\mathbf{v}}'}{2}\right>,
\end{eqnarray}
where the symbol
$\left<...\right>$
stands for the angular average.
The disorder correlations between the layers are described by 
the characteristic rate
\begin{equation}
\frac{1}{\tau_g}=\frac{\tau_{12}-\tau}{\tau^2},
\label{taug}
\end{equation}
where $1/\tau = [1/ \tau_{11}+ 1/ \tau_{22}]/2$.
The time scale $\tau_g$ is a characteristic scale on which carriers
in the two layers start ``feeling'' the difference between the impurity
potentials $u^{(1)}$ and $u^{(2)}$.
The potentials in the two layers are strongly correlated 
when $\tau_g\gg \tau$. One might expect that for realistic systems
the situation of moderately correlated potentials,
$\tau_g\sim \tau \sim \tau_{12},$ is typically realized.
Weakly correlated potentials ($\tau_{12}\gg \tau$) yield $\tau_g\ll\tau$.
Below we assume that disorder is sufficiently
short-ranged and do not distinguish between the total 
and transport scattering rates for the estimates.

We start from the ballistic regime $T\tau\gg 1$. 
The correlated disorder affects the drag in a way
similar to the third-order drag.
With correlated disorder, one can include an interlayer disorder line $w_{12}$
into the inelastic scattering amplitude.
In the ballistic $\rho^{(3)}_D$ drag we had one amplitude $M_2$ with two
interaction lines and one with a single wave line ($M_1$).
The corresponding drag rate contains $2Re[M_1(M_2)^*] \propto \alpha^3$.
Now one can form the second-order scattering amplitude $M_2$ using one 
interaction line ($\alpha$) and one interlayer-disorder line, which introduces a
factor $(T\tau_{12})^{-1}$. This gives 
\begin{equation}
\frac{1}{\tau_D^{\text{corr}}} \sim \alpha^2 T (T \tau_{12})^{-1}=\alpha^2/\tau_{12},
\end{equation}  
and
\begin{equation}
\rho_D^{\text{corr}}\sim \frac{\alpha^2}{T \tau_{12}},
\label{rhocorrball}
\end{equation}
which overcomes the third-order drag 
$\rho_D^{(3)} \sim \alpha^3$ for $1/\tau_{12}>\alpha T$. This happens in the
perturbative regime ($1/\tau>\alpha^2 T$, assuming correlated disorder, $\tau_{12}\sim \tau$), where
the correlated-disorder contribution can be calculated diagrammatically.
Similarly to $\sigma_D^{(3)}$, the corresponding diagram involves
two four-leg vertices (hence finite drag at the Dirac point $\mu=0$), but now connected
in all possible ways by two interaction lines and one disorder line $w_{12}$.

The general expression for the drag resistivity in the ballistic regime,
including both third-order and correlated-disorder drag rates for equal layers has the form:
\begin{equation}
\rho_D\sim \frac{\hbar}{e^2}\frac{1}{NT}
\dfrac{\dfrac{N}{\tau}\left(\alpha^2 \dfrac{\mu^2}{T^2} + \alpha^3 T + \dfrac{\alpha^2}{T\tau_{12}}\right)+\dfrac{\mu^2 N^2}{T^2}\left[\alpha^4 T^2-\left(\alpha^2 \dfrac{\mu^2}{T^2} + \alpha^3 T + \dfrac{\alpha^2}{T\tau_{12}}\right)^2 \right]}
{\dfrac{1}{\tau}+\dfrac{\mu^2 N}{T^2} \left[\alpha^2 T + \left(\alpha^2 \dfrac{\mu^2}{T^2}  + \alpha^3 T+ \dfrac{\alpha^2}{T\tau_{12}}\right) \right]}
\end{equation}
Exactly at the Dirac point it reduces to:
 \begin{equation}
\rho_D(\mu=0)\sim \frac{\hbar}{e^2}\alpha^2\left(\frac{1}{T\tau_{12}}+\alpha\right).
\end{equation}

Let us now analyze the role of correlated disorder in the diffusive regime $T\tau\ll 1$.
Again, we assume the absence of localization at the Dirac point (see Section 3).
The drag resistivity for the case of correlated disorder was calculated in the diffusive regime
in Ref. \cite{gor}. It is dominated by the Maki-Thompson diagram with an interlayer Cooper propagator.
It is worth noting that any difference in the disorder potentials (as well as in chemical potentials of the layers)
leads to a finite gap in these propagators given by $1/\tau_g$.
The main result of Ref. \cite{gor} is as follows: 
\begin{equation}
\rho^{\text{corr}}_D\sim 
\frac{\hbar}{e^2} \frac{1}{g^2
[\lambda^{-1}_{21} + \ln(\varepsilon_0 /T)]^2}
\ln\frac{T\tau_{\varphi}\tau_{g}}
{\tau_{\varphi}+\tau_{g}}
\label{rhototal1}
\end{equation}
at $\tau^{-1}_{g} \ll T \ll \tau^{-1}$, and
\begin{equation}
\rho^{\text{corr}}_D\sim 
\frac{\hbar}{e^2} \frac{(T\tau_g)^2}{g^2
[\lambda^{-1}_{21} + \ln(\varepsilon_0 \tau_g)]^2}.
\label{rhototal2}
\end{equation}
at lower temperatures $T\ll \tau^{-1}_{g}$.

In graphene near the Dirac point, for small interlayer distance $\kappa d\ll 1$ 
the interlayer interaction constant in the Cooper channel is $\lambda_{12}\sim \alpha$. 
The Cooper channel cutoff energy is $\epsilon_0=1/\tau$
(the logarithm in the Cooper channel appears only for a constant density
of states; in graphene in the diffusive regime this happens only for energies below $1/\tau$),
the dimensionless conductance $g \sim 1$, and $\tau_\phi \sim 1/T$.
Substituting these values to Eqs. (\ref{rhototal1}) and (\ref{rhototal2}), we arrive at
\begin{eqnarray}
\rho^{\text{corr}}_D &\sim& \frac{\hbar}{e^2} \frac{\alpha^2 }{[1-\alpha \ln(T \tau)]^2},\qquad \tau^{-1}_{g} \ll T \ll \tau^{-1},
\label{rhocorr1}
\\
&&
\nonumber
\\
\rho^{\text{corr}}_D &\sim& \frac{\hbar}{e^2} \frac{\alpha^2 (T \tau_g)^2}{[1-\alpha \ln(T \tau)]^2}, \qquad T\ll \tau^{-1}_{g}.
\label{rhocorr2}
\end{eqnarray}
These results are $\propto \alpha^2$ for realistic temperatures,
$T \tau \gg \exp(-1/\alpha)$.
For a moderately correlated disorder $\tau_g\sim \tau$, Eqs.~(\ref{rhocorrball}) and (\ref{rhocorr2}) then lead to
\begin{equation}
\rho^{\text{corr}}_D \sim \frac{\hbar}{e^2} \alpha^2 \begin{cases} (T\tau)^{-1}, &\qquad T\tau\gg 1 \\[0.5pt]
(T\tau)^2, &\qquad T\tau\ll 1 \end{cases},
\end{equation} 
which yields a maximum at $T\sim 1/\tau$ in the temperature dependence of the drag resistivity at the
charge neutrality point. For strongly correlated disorder potentials ($\tau_g\gg \tau$), this maximum develops into a plateau
between $\tau^{-1}_{g} \ll T \ll \tau^{-1}$.

\subsection{B. Long-range correlations: correlated macroscopic inhomogeneities}

Let us now analyze within our kinetic-equation framework
the model of correlated macroscopic spatial fluctuations $\delta \mu_i$ in chemical potentials
of the two layers \cite{lev}, characterized by the correlation function
\begin{equation}
F_{ij}^{(\mu)}(\bs{r}-\bs{r}')=\langle\delta\mu_i(\bs{r})\delta\mu_j(\bs{r}')\rangle\neq 0.
\end{equation}
We restrict ourselves to the ballistic regime $T\tau\gg 1$.
Assuming the spatial scale of the fluctuations to be much larger than all characteristic
scales related to the particle scattering, $v\tau_{ee}$, $v\tau_D$, and $v\tau$, we solve the 
hydrodynamic equations locally, yielding Eq.~(\ref{rhoab}) with local values of the chemical potentials
encoded in functions $B_1^{a(b)}\sim \mu_{a(b)}/T$, as well as in the local drag rate
$$
\frac{1}{\tau_D(\bs{r})}\sim \alpha^2 N \frac{\mu_1(\bs{r})\mu_2(\bs{r})}{T}.
$$
On the other hand, since the coefficients $B_0\sim 1$ and $B_2\sim 1$, as well as the transport electron-electron
scattering rate $\tau_{ee}^{-1}\sim \alpha^2 T$ are finite at the neutrality point,
we can neglect the fluctuations of $\mu_{i}$ in these quantities.
Exactly at the Dirac point $\mu_{1,2}=0$, assuming that the fluctuations of chemical potentials are weak (the precise condition
is established below),
we can further neglect the $B_1$-terms in the denominator of Eq.~(\ref{rhoab}), yielding 
for the ``local resistivity''
\begin{eqnarray}
\rho_D(\bs{r})\simeq\frac{\hbar}{e^2}\frac{2 B_2 \tau}{N  B_0^a B_0^b T}
\left[\dfrac{1}{\tau\tau_D(\bs{r})}+\dfrac{B_1^a(\bs{r})B_1^b(\bs{r})}{\tau^a_{ee}\tau^b_{ee}}\right]
\nonumber \\
&&
\nonumber \\
\sim \frac{\hbar}{e^2}\frac{\tau}{N T}\, \delta \mu_1(\bs{r}) \delta\mu_2(\bs{r})\,
\left(\dfrac{\alpha^2 N}{T \tau}+\alpha^4 N^2\right)
.
\label{rhoabcorr}
\end{eqnarray}
Averaging this expression over the small fluctuations of the
correlated chemical potentials \cite{lev}, we arrive at the
correction to the universal third-order result,
$\rho_D^{(3)}(\mu=0)\sim (\hbar/e^2)\alpha^3$,
\begin{equation}
\Delta \rho_D(\mu=0) \sim \frac{\hbar}{e^2}\frac{\alpha^2\, F^{(\mu)}_{12}(0)}{T^2}\, \left(1+\alpha^2 N T\tau\right)
\sim \frac{\hbar}{e^2}\frac{F^{(\mu)}_{12}(0)}{T^2}\,
\begin{cases}
\alpha^4 N T\tau &\qquad \dfrac{1}{\tau}\ll \alpha^2 N T,\\[0.2pt]
\alpha^2 &\qquad \alpha^2 N T\ll \dfrac{1}{\tau}\ll T.
\end{cases}
\end{equation}
We see that in the Coulomb-dominated transport regime, this correction
is dominated by the fluctuations in $B_1$, whereas in the disorder-dominated (perturbative) regime, the main role
is played by a locally finite drag rate.

Finally, in the ultraclean limit 
\begin{equation}
\frac{1}{\tau}\ll \alpha^2 N F_{ii}^{(0)}/T,
\label{cond-delta-mu}
\end{equation}
we can neglect $1/\tau$ in the denominator of the local drag resistivity given by Eq.~(\ref{rhoab}),
yielding a natural analog of Eq.~(\ref{r1dp}):
\begin{equation}
\Delta \rho_D(\mu=0)(\mathbf{r}) \sim \frac{\hbar}{e^2}\alpha^2 \frac{\delta\mu_1\delta\mu_2}{\delta\mu_1\delta\mu_1+\delta\mu_2\delta\mu_2}.
\label{DeltaRho}
\end{equation}
In particular, for perfectly correlated chemical potentials, $\delta\mu_1(\mathbf{r})=\delta\mu_2(\mathbf{r})$, the fluctuations drops out from 
Eq.~(\ref{DeltaRho}) and the local resistivity turns out to be independent of $\mathbf{r}$. 
In a more general case, the averaging over fluctuations becomes nontrivial, but this can only affect the numerical prefactor in the
final result. 
Thus, the correlated large-scale fluctuations of the chemical potentials in the layers in effect shift the curve 1 in Fig. \ref{pd} upwards,
extending the validity of the fully equilibrated result,
\begin{equation}
\rho_D \sim  \frac{\hbar}{e^2}\alpha^2,
\end{equation}
to the case of finite disorder, Eq.~(\ref{cond-delta-mu}), at the Dirac point. 
This implies that in the case of correlated inhomogeneities the disorder-induced dip in the lower 
left panel of Fig. 1 develops only for sufficiently strong disorder.

\bigskip

\end{widetext}


\begin{thebibliography}{99}
\bibitem{fn2} 
We treat impurity scattering within the relaxation time approximation
assuming energy-independent transport time $\tau$. Generalization onto
the more realistic case of Coulomb impurities with the
energy-dependent transport time is straightforward and does not
qualitatively affect the results.
\end{thebibliography}
\end{document}